\documentclass[prd,nofootinbib,onecolumn,preprint,11pt]{revtex4}

\usepackage{graphicx}
\usepackage{bm}
\usepackage{float}
\usepackage{subfigure}
\usepackage{amsmath}
\usepackage{amsfonts}
\usepackage{color}
\usepackage{slashed}
\usepackage{fancyhdr}
\usepackage[colorlinks=true,linkcolor=black,citecolor=black,filecolor=black,urlcolor=black,unicode]{hyperref}

\newcommand{\GN}{G_{\rm N}}

\newcommand{\dd}{\text{d}}

\begin{document}

\title{Gravitational Tensor and Acoustic Waves in A Radiation Dominated Universe: Weyl Curvature and Polarization Patterns}

\author{Yi-Zen Chu$^{1,2}$ and Yen-Wei Liu$^1$}
\affiliation{
$\,^1$Department of Physics, National Central University, Chungli 32001, Taiwan \\
$\,^2$Center for High Energy and High Field Physics (CHiP), National Central University, Chungli 32001, Taiwan
}

\begin{abstract}
\noindent We provide evidence that gravitational radiation in a 4D radiation-dominated universe, with equation-of-state $w=1/3$, consists of two components: helicity-2 gravitons and massless scalar acoustic waves. On physical grounds, we would expect the homogeneous solution of the Weyl tensor components to be a good approximation to its inhomogeneous counterparts, whenever the observer is located well in the far zone of an isolated astrophysical source of cosmological gravitational radiation. We show explicitly that these homogeneous and inhomogeneous solutions both receive contributions from the gauge-invariant tensor {\it and} the two Bardeen (acoustic) scalars. Comparison of these Weyl tensor computations thus allows us to not only identify, in the high frequency limit, the corresponding gravitational tensor and scalar radiation; but also their oscillatory polarization patterns.
\end{abstract}

\maketitle

\tableofcontents

\section{Motivation}
\label{Section_Introduction}

What is gravitational radiation? This question is well studied in the asymptotically flat case, where the metric takes the form $g_{\mu\nu} \to \eta_{\mu\nu} + h_{\mu\nu}$. The asymptotically cosmological case is much less studied at a fundamental theoretical level, however. For instance, basic questions such as `What is the gravitational-wave energy produced by an isolated source in a cosmological background?' have only begun to be answered in the specific case of de Sitter/Cosmological Constant dominated universe in the very recent works of \cite{Ashtekar:2015lxa}, \cite{Date:2016uzr}, \cite{Bonga:2017dlx}, and \cite{Hoque:2017xop}, although the generic JWKB results are known in the high frequency limit \cite{Isaacson:1967zz,Isaacson:1968zza}.

In this paper, we wish to initiate an investigation into whether there are additional radiative degrees of freedom associated with the metric when the cosmology is driven by a perfect fluid -- as opposed to, say, the vacuum spacetimes of Minkowski and de Sitter -- and if so, what the corresponding oscillatory polarization patterns are. More specifically, we place a hypothetical isolated astrophysical system in an even $(d \geq 4)-$dimensional universe, which in turn is assumed to be driven by a single perfect fluid of constant equation-of-state $w$; and we inquire what sort of gravitational waves it engenders.\footnote{Our primary goal here is to obtain the concrete cosmological GW signals received from a remote isolated source. For a rigorous formal treatment of nonlinear GWs, we refer the reader to the book \cite{Choquet-Bruhat:2009xil}.  \\	
The weak cosmological GWs were also studied by Grischuk \cite{Grishchuk:1974ny}; however, he only considered source-free metric perturbations and not the fluid equations themselves. After imposing a gauge condition and specializing to the cosmological background, he was essentially dealing with the spin-2 portion of the metric perturbations, in contrast to our attempt to solve for the full metric perturbations in the Einstein-fluid system (note that his eq.~(8) corresponds to the homogeneous version of our eq.~\eqref{Spin-2_WaveEquation_Cosmology}). On the other hand, the focus of his work is on the low-frequency regime where the GWs would be amplified as long as the universe is not in its radiation-dominated phase, whereas we are interested in the high-frequency GWs produced by the distant astrophysical sources (see \S \eqref{PolarizationPattern_4DRadiation} below for the far-zone JWKB analysis).} Hence, our perturbed cosmological geometry, containing this gravitational radiation, takes the form\footnote{The ``mostly plus'' sign convention is used for the metric, namely $\eta_{\mu\nu} = \text{diag}[-1,+1,\ldots,+1]$; Greek indices $\mu, \nu,\ldots,$ run from $0$ (time) to $d-1$, while Latin ones $i, j, \ldots,$ run over only the spatial values $1$ to $d-1$. The symmetrization/anti-symmetrization of tensor indices are denoted by round $(\ldots)$/square $[\ldots]$ brackets, e.g., $T_{(\mu\nu)} \equiv \frac12 \left( T_{\mu\nu} + T_{\nu\mu}\right)$ and $T_{[\mu\nu]} \equiv \frac12 \left( T_{\mu\nu}  -  T_{\nu\mu} \right)$.}
\begin{align}
\label{PerturbedFLRW}
g_{\mu\nu}[\eta,\vec{x}] &= a[\eta]^2 \left( \eta_{\mu\nu}+\chi_{\mu\nu}[\eta,\vec{x}] \right) ; \\
\label{ScaleFactor}
a[\eta] &= \left(\frac{\eta}{\eta_0}\right)^{\frac{2}{q_w}}, \qquad
q_w \equiv (d-3) + (d-1) w , \qquad
w=-1,\ 0 \leq w \leq 1 .
\end{align}
For the discussion at hand, it is advantageous to work in the synchronous gauge $\chi^{(s)}_{\mu 0} = 0 $ because the proper geodesic spatial distance between two co-moving freely falling test masses at $\vec{x}=\vec{Y}_0$ and $\vec{x}=\vec{Z}_0$ may then be readily expressed as
\begin{align}
L\left[\eta;\vec{Y}_0 \leftrightarrow \vec{Z}_0\right]
&= a[\eta] \big| \vec{Y}_0 - \vec{Z}_0 \big| \left( 1 + \frac{1}{2} \widehat{n}^i \widehat{n}^j \int_{0}^{1} \chi^{(s)}_{ij}\left[\eta,\vec{Y}_0 + \lambda(\vec{Z}_0-\vec{Y}_0)\right] \dd \lambda + \mathcal{O}\left[ \big(\chi^{(s)}_{ab}\big)^2\right] \right) ; \\
\widehat{n}^i
&\equiv \frac{Y_0^i - Z_0^i}{\big|\vec{Y}_0 - \vec{Z}_0\big|} .
\end{align}
We see the fractional distortion of space is, at leading order in $L (\dot{a}/a)$,
\begin{align}
\label{FractionalDistortion}
\left(\frac{\delta L}{L_0}\right)\left[\eta;\vec{Y}_0 \leftrightarrow \vec{Z}_0\right]
&=  \frac{1}{2} \widehat{n}^i \widehat{n}^j \int_{0}^{1} \chi^{(s)}_{ij}\left[\eta,\vec{Y}_0 + \lambda(\vec{Z}_0-\vec{Y}_0)\right] \dd \lambda + \mathcal{O}\left[\big(\chi^{(s)}_{ab}\big)^2\right] .
\end{align}
Now, gravitational radiation itself may be characterized by its ability to do work on a hypothetical Weber bar immersed in a curved spacetime, due to tidal forces induced by the finite-frequencies content of geometric curvature -- namely, the time dependent portion of the geodesic deviation equation tells us the relative acceleration $a^i$ between infinitesimally nearby test masses is
\begin{align}
\label{TidalForces}
a^i \equiv \nabla_U \nabla_U \xi^i = - B^i_{\phantom{i}j} \xi^j .
\end{align}
Here $\nabla_U$ denotes the covariant derivative along the co-moving family of timelike observers in cosmology, with tangent vector $U^\mu = a^{-1} \delta_0^\mu$ in the synchronous gauge. Moreover, within the same gauge, the exact distortion tensor is
\begin{align}
B^i_{\phantom{i}j}
\equiv R^i_{\phantom{i} \alpha j \beta} U^\alpha U^\beta
= a^{-2} R^i_{\phantom{i} 0 j 0}.
\end{align}
We shall focus on the traceless part of this tidal tensor -- i.e., the Weyl components $C^i_{\phantom{i}0j0}$.
\begin{align}
a^i
= - \left( (\text{trace-terms})^i_{\phantom{i} 0 j 0} + a^{-2} \delta_1 C^i_{\phantom{i} 0 j 0} + \mathcal{O}\left[\big(\chi^{(s)}_{ab}\big)^2\right] \right) \xi^j  .
\end{align}
The reasons are two-fold. The first is geometric: the traceless part of the tensor $B^i_{\phantom{i}j}$ produces an independent distortion pattern from its trace part; see \cite{Poisson:2009pwt} for a pedagogical discussion. The second is technical: because the Weyl tensor $C^\alpha_{\phantom{\alpha}\beta \mu\nu}$ is conformally invariant, it is zero when there are no perturbations, and hence its linear-in-$\chi_{\mu\nu}$ piece $\delta_1 C^i_{\phantom{i} 0 j 0}$ is gauge-invariant. We may therefore use the gauge-invariant content of $\chi_{\mu\nu}$ obtained in \cite{Chu:2016ngc} to construct the solution for $\delta_1 C^i_{\phantom{i}0j0}$; and in turn, the high-frequency synchronous gauge perturbation $\chi^{(s)}_{ij}$. Note that the computation of $\delta_1 C^i_{\phantom{i}0j0}$ has already been partially undertaken in \cite{Chu:2019ndv}; but in this work we will not only provide somewhat simpler final results but also explicit solutions in the physically important case of the four-dimensional (4D) radiation dominated universe (i.e., $d=4$ and $w = 1/3$). The 4D solutions for $\chi^{(s)}_{ij}$ during radiation domination will teach us; at least in the short wavelength limit, isolated gravitational-wave (GW) sources engender 2 independent sets of oscillatory polarization degrees-of-freedom. These two sets of polarizations survive even when the gravitational-wave source is removed; that is, both the inhomogeneous and homogeneous solutions of $\delta_1 C^i_{\phantom{i}0j0}$ contain them. Since the latter may be viewed as an approximation to the situation where the observer is located in the distant far zone of the source, our results therefore suggest that gravitational radiation in a 4D radiation universe indeed does not consist solely of the usual massless spin-2 modes but also of the Bardeen scalars, because both the tensor and scalar gauge invariant variables appear in the homogeneous solutions of $\delta_1 C^i{}_{0j0}$.\footnote{The primary goal of this paper is to study the propagation of cosmological GWs in the position space, in contrast to the Fourier space analysis that cosmologists usually perform to compute the large-scale structure correlation functions.}

In \S \eqref{LinearizedGravity_RelativisticFluid}, we will gather and improve upon the results from \cite{Chu:2016ngc} and \cite{Chu:2019ndv}. We delineate both the gauge-invariant content of $\chi_{\mu\nu}$ and the general construction of the Weyl tensor that was begun in \cite{Chu:2019ndv}. In \S \eqref{PolarizationPattern_4DRadiation}, we specialize to the primary cosmology of physical interest, the $d=4$ and $w=1/3$ case. Then in \S \eqref{Section_Summary} we summarize the results and sketch future directions. In appendix \eqref{LinearizedWeyl_deSitterAndMatterDomination}, we tie up some loose ends from \cite{Chu:2019ndv}; and describe the linearized Weyl tensor $\delta_1 C^i_{\phantom{i}0j0}$ results in de Sitter $w=-1$ and matter dominated $w=0$ universes.

\section{Linearized Gravitation in Spatially-Flat Cosmologies with Relativistic Fluid} \label{LinearizedGravity_RelativisticFluid}

{\bf Gauge-invariant variables} \qquad In \cite{Chu:2016ngc} and \cite{Chu:2019ndv}, the following scalar-vector-tensor decomposition was performed for the metric perturbation $\chi_{\mu\nu}$ of eq.~\eqref{PerturbedFLRW} as well as the stress-energy tensor ${}^{(\text{a})}T_{\mu\nu}$ of the hypothetical isolated astrophysical system.
\begin{align}
\label{Bardeen_SVTDecomposition}
\chi_{00} &\equiv E, \qquad \qquad \chi_{0i} \equiv \partial_i F + F_i, \notag\\
\chi_{ij} &\equiv D_{ij} + \partial_{( i} D_{j) } + \frac{D}{d-1} \delta_{ij} + \left( \partial_i \partial_j - \frac{\delta_{ij}}{d-1} \vec{\nabla}^2 \right) K ,
\end{align}
\begin{align}
\,^{(\text{a})}T_{00} & \equiv \rho, \qquad\qquad
\,^{(\text{a})}T_{0i} \equiv \Sigma_i + \partial_i \Sigma, \notag\\
\label{Astro_SVT}
\,^{(\text{a})}T_{ij} & \equiv \sigma_{ij} + \partial_{ (i } \sigma_{ j ) } + \frac{\sigma}{d-1} \delta_{ij}
+ \left( \partial_i \partial_j - \frac{\delta_{ij}}{d-1} \vec{\nabla}^2 \right) \Upsilon .
\end{align}
By construction, these modes are subject to the following constraints:
\begin{align}
\label{Bardeen_SVTConstraints}
\partial_i F_i &= \partial_i D_i = 0 = \delta^{ij} D_{ij} = \partial_i D_{ij} , \\
\label{AstroTmunu_SVTConstraints}
\partial_i \Sigma_i &= \partial_i \sigma_i = 0 = \delta^{ij} \sigma_{ij} = \partial_i \sigma_{ij} .
\end{align}
From eq.~\eqref{Bardeen_SVTDecomposition}, the gauge-invariant content of $\chi_{\mu\nu}$ takes the form of 2 Bardeen scalars $\Phi$ and $\Psi$; 1 vector $V_i$; and 1 spin-2 tensor $D_{ij}$; the following definitions ensure they are not altered under an infinitesimal change in coordinates.\footnote{The notations defined here are the same ones used in \cite{Chu:2019ndv}. To convert them into those defined in \cite{Chu:2016ngc}, we follow the replacements: $\Phi[\text{here}] \to \Psi[\text{\cite{Chu:2016ngc}}]/2$, $\Psi[\text{here}] \to \Phi[\text{\cite{Chu:2016ngc}}]/2$, $V_i[\text{here}] \to -V_i[\text{\cite{Chu:2016ngc}}]$, and $D_{ij}[\text{here}] \to - D_{ij}[\text{\cite{Chu:2016ngc}}]$.}
{\allowdisplaybreaks
\begin{align}
\label{Bardeen_Psi}
\Phi &\equiv -\frac E2 + \frac{1}{a} \partial_0 \left\{ a \left( F - \frac{\dot{K}}{2} \right) \right\} , \\
\label{Bardeen_Phi}
\Psi &\equiv -\frac{D - \vec{\nabla}^2 K}{2(d-1)} -  \frac{\dot{a}}{a} \left( F - \frac{\dot{K}}{2} \right) , \\
\label{Bardeen_VandDij}
V_i &\equiv F_i - \frac{\dot{D}_i}2
\qquad \qquad \text{ and }\qquad \qquad
D_{ij} \equiv \chi_{ij}^\text{TT} .
\end{align}}
{\bf Linearized Einstein equations} \qquad In \cite{Chu:2016ngc}, the gauge invariant variables were shown to obey
{\allowdisplaybreaks
	\begin{align}
	-\ddot\Psi-\big(q_w+d-2\big)\mathcal H\dot\Psi+w\vec\nabla^2\Psi
	& = -8\pi G_\text N\bigg(\frac{\partial_0\left(a^{d-2}\Sigma\right)}{(d-2)a^{d-2}}-\frac{w\rho}{(d-2)}+\mathcal H\dot{\Upsilon}\bigg),
	\label{Psi_WaveEq_Cosmology} \\
	(d-3) \Psi - \Phi & = 8\pi \GN \Upsilon ,
	\label{PhiPsiRelationship} \\
	\vec\nabla^2 V_i&=-16\pi G_\text N \Sigma_i,
	\label{Vi_Poisson_Cosmology}\\
	-\ddot D_{ij}-(d-2)\mathcal H\dot D_{ij}+\vec\nabla^2 D_{ij}
	& = -16\pi G_\text N\sigma_{ij} ,
	\label{Spin-2_WaveEquation_Cosmology}
	\end{align}}%
where $\mathcal H \equiv \dot a / a = 2 /(q_w \eta)$ denotes the conformal Hubble parameter. These field equations were first solved analytically in \cite{Chu:2016ngc}, written in terms of their associated scalar Green's functions convolved against the non-local components of the matter stress-energy tensor\footnote{Other than the energy density $\rho$ and the spatial trace $\sigma$, the rest of the components in eq.~\eqref{Astro_SVT} are the non-local functions of the original stress-energy tensor $^{(\text a)}T_{\mu\nu}$. This non-locality arises from the local projections implemented in Fourier space that leads up to the weighted smearing of $^{(\text a)}T_{\mu\nu}$ all over the space.}, namely the right-hand side of eqs.~\eqref{Psi_WaveEq_Cosmology}-\eqref{Spin-2_WaveEquation_Cosmology}. Then, with the help of the Fourier-space projection and the ``time-integral'' method developed in \cite{Chu:2019ndv}, these solutions can be further re-cast into the convolutions of their effective Green's functions with the local components of the astrophysical $^{(\text a)}T_{\mu\nu}$.

{\it Spin-2} \qquad The solution to eq.~\eqref{Spin-2_WaveEquation_Cosmology}, governing the spin-2 graviton $D_{ij}$, can be found in eq.~(247) of \cite{Chu:2019ndv}:
{\allowdisplaybreaks
	\begin{align}
	D_{ij}[\eta,\vec x]  & =   16\pi G_\text N\int_{\mathbb{R}^{d-1}} \dd^{d-1}\vec x'\int_0^\infty \dd\eta'\, \Theta[T] \left(\frac{a[\eta']}{a[\eta]}\right)^{\frac {d-2}2}\Bigg\{ C^{(g)}_{1,d}\bigg({}^{(\text{a})}T_{ij}[\eta',\vec x'] \notag\\
	& + \frac{\delta_{ij}}{d-2}\left( {}^{(\text{a})}T_{00}[\eta',\vec x'] - {}^{(\text{a})}T_{ll}[\eta',\vec x']  \right)\bigg)  - 2a[\eta']^{\frac{d-2}2}\partial_{\eta'}\left(a[\eta']^{-\frac{d-2}2}\partial_{(i}C^{(g)}_{2,d}\right){}^{(\text{a})}T_{j)0}[\eta',\vec x']  \notag\\
	& + \frac{\delta_{ij}}{d-2}  \mathcal H[\eta']  a[\eta']^{\frac{d-2}2}\,\partial_{\eta'}\left(a[\eta']^{-\frac{d-2}2}C^{(g)}_{2,d}\right)     \left((d-3){}^{(\text{a})}T_{00}[\eta',\vec x']+{}^{(\text{a})}T_{ll}[\eta',\vec x']\right) \notag\\
	&  - \frac1{d-2} \bigg( \partial_i\partial_jC^{(g)}_{2,d} - (d-3)\mathcal H[\eta'] a[\eta']^{\frac{d-2}2}\partial_{\eta'}\left(a[\eta']^{-\frac{d-2}2}\partial_i\partial_jC^{(g)}_{3,d}\right) \bigg)
	\left( (d-3) {}^{(\text{a})}T_{00}[\eta',\vec x']  +  {}^{(\text{a})}T_{ll}[\eta',\vec x']\right)    \Bigg\}  \notag\\
	& +\frac{16\pi G_\text N}{d-2}\int_{\mathbb{R}^{d-1}} \dd^{d-1}\vec x'  \left(\delta_{ij}G^{\mathrm{(E)}}_d{} ^{(\text{a})}T_{00}[\eta,\vec x']  +(d-3)\partial_i\partial_jD_d{}^{(\text{a})}T_{00}[\eta,\vec x']\right),
	\label{Spin2Solution_RadiationDominated}
	\end{align}}%
where $T \equiv \eta - \eta'$, $R\equiv|\vec x - \vec x'|$, and $^{(\text{a})}T_{ll}\equiv\delta^{ij}\,^{(\text{a})}T_{ij}$. The scalar function $C^{(g)}_{1,d}$ itself, obeying a homogeneous wave equation, is the advanced minus retarded scalar Green's function,
\begin{align}
C^{(g)}_{1,d}[\eta,\eta';R] = G^{(g,-)}_d[\eta,\eta';R] - G^{(g,+)}_d[\eta,\eta';R]
\label{Commutator_Spin2} .
\end{align}
Equivalently, $G^{(g,+)}_d = -\Theta[T]C^{(g)}_{1,d}$. For the even dimensional case,
\begin{align}
\label{GeneralSol_1stGreensFunction_EvenDimensions}
G^{(g,+)}_{\text{even }d\geq4}[\eta,\eta';R]
&  = - \Theta[T] \left(  \frac1{2\pi} \frac{\partial}{\partial \overline \sigma} \right)^{\frac{d-2}2} \left( \frac{\Theta[\overline\sigma]}2 P_{ - \frac{d-2}{q_w} }\left[1+\frac{\overline\sigma}{\eta\eta'}\right]\right) , \qquad \overline\sigma = \frac{(\eta - \eta')^2 - R^2}2  ,
\end{align}
where $P_{\nu}$ is the Legendre function. (See also eqs.~(112) and (113) of \cite{Chu:2016ngc}.) Note that all the scalar Green's functions introduced in this paper, generally denoted by $G^+_d$, obey the following type of wave equation:
\begin{align}
\left(  \partial^2_{\eta,\vec x}  + \frac{ \kappa(\kappa + 1)}{\eta^2} \right) G^+_d = \left(  \partial^2_{\eta',\vec x'}  + \frac{ \kappa(\kappa + 1)}{\eta'^2} \right) G^+_d  = \delta[\eta - \eta'] \delta^{(d - 1)}[\vec x - \vec x'],
\label{WaveEq_ScalarGreensFunction}
\end{align}
where $\partial^2 \equiv \eta^{\mu\nu} \partial_\mu \partial_\nu $. Here, $G^{(g,+)}_d$ solves eq.~\eqref{WaveEq_ScalarGreensFunction} with $\kappa = - (d - 2)/ q_w$. Additionally, $C^{(g)}_{2,d}$ and $C^{(g)}_{3,d}$ are both related to the known $C^{(g)}_{1,d}$ through the ``time-integral method'' developed in \cite{Chu:2019ndv}:
{\allowdisplaybreaks
	\begin{align}
	C^{(g)}_{2,d}[\eta,\eta';R]=\,&-a[\eta]^{\frac {d-2}2}\int^\eta_{\eta'}\dd\eta_2\,a[\eta_2]^{-(d-2)}\int^{\eta_2}_{\eta'}\dd\eta_1\,a[\eta_1]^{\frac {d-2}2}C^{(g)}_{1,d}[\eta_1,\eta';R] \notag\\
	&-G^{(\mathrm E)}_d[R]\left(\frac{a[\eta]}{a[\eta']}\right)^{\frac {d-2}2}\int^\eta_{\eta'}\dd\eta_1\left(\frac{a[\eta']}{a[\eta_1]}\right)^{d-2},
	\label{C2_Graviton_TimeIntegral_Cosmology} \\
	C^{(g)}_{3,d}[\eta,\eta';R]=\,&a[\eta]^{\frac {d-2}2}\int^\eta_{\eta'}\dd\eta_4\,a[\eta_4]^{-(d-2)}\int^{\eta_4}_{\eta'}\dd\eta_3\,a[\eta_3]^{d-2}\int^{\eta_3}_{\eta'}\dd\eta_2
	\,a[\eta_2]^{-(d-2)}\int^{\eta_2}_{\eta'}\dd\eta_1\,a[\eta_1]^{\frac {d-2}2}C^{(g)}_{1,d}[\eta_1,\eta';R] \notag\\
	&+G^{(\mathrm E)}_d[R]\left(\frac{a[\eta]}{a[\eta']}\right)^{\frac{d-2}2}\int^\eta_{\eta'}\dd\eta_3\,a[\eta_3]^{-(d-2)}\int^{\eta_3}_{\eta'}\dd\eta_2
	\,a[\eta_2]^{d-2}\int^{\eta_2}_{\eta'}\dd\eta_1\left(\frac{a[\eta']}{a[\eta_1]}\right)^{d-2} \notag\\
	&+D_d[R]\left(\frac{a[\eta]}{a[\eta']}\right)^{\frac {d-2}2}\int^\eta_{\eta'}\dd\eta_1\left(\frac{a[\eta']}{a[\eta_1]}\right)^{d-2},
	\label{C3_Graviton_TimeIntegral_Cosmology}
	\end{align}}%
with the Euclidean Green's function $G^{(\mathrm E)}_d$ and $D_d$, respectively, defined as the inverse Fourier transforms of $-1/\vec k^2$ and $1/\vec k^4$,
{\allowdisplaybreaks
	\begin{align}
	G^{(\text{E})}_{d\geq4}[R]
	& = - \frac{\Gamma[\frac{d-3}{2}]}{4\pi^{\frac{d-1}{2}} R^{d-3}} ,
	\label{EuclideanGreenFunction} \\
	D^{(\text{reg})}_{4}[R] & =  -\frac{R}{8\pi}      ,
	\label{D4}  \\
	D^{(\text{reg})}_{5+2\epsilon}[R] & =   \frac1{16\pi^2}  \left( \frac1\epsilon -\gamma - \ln[\pi] - 2\ln[\mu R] \right)    ,
	\label{D5}   \\
	D_{d\geq6}[R] & = \frac{\Gamma\left[\frac{d-5}2\right]}{16\pi^{\frac{d-1}{2}}R^{d-5}} ;
	\label{Dd_Expression}
	\end{align}}%
note that $D_4$ and $D_5$ of the latter have been dimensionally regularized, in which an arbitrary mass scale $\mu$, as well as the Euler-Mascheroni constant $\gamma$, were introduced, but those constants present in the regularization scheme will not show up in the final results because they will be removed by the spatial derivatives $\partial_i\partial_j$ of the tensor structure in eq.~\eqref{Spin2Solution_RadiationDominated}. Note that the expression \eqref{Spin2Solution_RadiationDominated} here is a bit different from eq.~(247) of \cite{Chu:2019ndv}, where the double-time derivatives of $C^{(g)}_{2,d}$ and $C^{(g)}_{3,d}$ with respect to $\eta'$ have been replaced with the lower-derivative terms via the homogeneous wave equation for $C^{(g)}_{1,d}$, which also holds for both $C^{(g)}_{2,d}$ and $C^{(g)}_{3,d}$.

{\it Vector Potential} \qquad As long as all the gravitational perturbations $\chi_{\mu\nu}$ become negligible in the far past, \cite{Chu:2016ngc} has argued that (cf.~eq.~(178) of \cite{Chu:2019ndv})
\begin{align}
V_i[\eta,\vec x]=16\pi G_\text N\int_{\mathbb{R}^{d-1}}   \dd^{d-1}\vec{x}'\left(\partial_i\partial_jD_d
{}^{(\text{a})}T_{0j}[\eta,\vec{x}']-G^{\mathrm{(E)}}_d{}^{(\text{a})}T_{0i}[\eta,\vec{x}']\right) .
\label{VectorMode_RadiationDominated}
\end{align}
{\it Bardeen Scalar Potentials for $0 < w \leq 1$} \qquad The inhomogeneous $\Psi$ solution that solves the scalar wave equation \eqref{Psi_WaveEq_Cosmology} was originally found in eq.~(264) of \cite{Chu:2019ndv} to involve all the three scalar functions $C^{(w)}_{1,d}$, $C^{(w)}_{2,d}$, and $C^{(w)}_{3,d}$ defined similarly to their spin-2 counterparts $C^{(g)}$s above. However, that expression can actually be further reduced to a more concise form involving just $C^{(w)}_{2,d}$ alone,
{\allowdisplaybreaks
	\begin{align}
	\hspace{-1em}\Psi[\eta,\vec x]=\,& - \frac{8\pi G_\text N}{d-2}  \frac{\left( 1 + \frac{q_w}2 \right)}{w^{\frac{d-3}2}}\int_{\mathbb{R}^{d-1}} \dd^{d-1}\vec x'\int_0^\infty \dd\eta' \, \Theta[T] \left(\frac{a[\eta']}{a[\eta]}\right)^{\frac{d-2}2}  \bigg(    \mathcal H[\eta] \mathcal H[\eta'] C^{(w)}_{2,d}  \left((d-3){}^{(\text{a})}T_{00}[\eta',\vec x'] \right. \notag\\
	& \left. + {}^{(\text{a})}T_{ll}[\eta',\vec x']\right) \bigg)  + \frac{8\pi G_\text N}{d-2}  \int_{\mathbb{R}^{d-1}}  \dd^{d-1}\vec x'\,\left( G^{(\mathrm E)}_d {}^{(\text{a})}T_{00}[\eta,\vec x']+(d-1)\mathcal H[\eta]\partial_jD_d {}^{(\text{a})}T_{0j}[\eta,\vec x']\right) .
	\label{BardeenPsi_RadiationDominated}
	\end{align}}%
The $C^{(w)}_{1,d}$ is the advanced minus retarded Green's function,
\begin{align}
C^{(w)}_{1,d}[\eta,\eta';R] = G^{(w,-)}_d[\eta,\eta';R] - G^{(w,+)}_d [\eta,\eta';R] ;
\label{Commutator_BardeenPsi}
\end{align}
while, in a similar manner to eq.~\eqref{C2_Graviton_TimeIntegral_Cosmology},  $C^{(w)}_{2,d}$ is expressible in terms of $C^{(w)}_{1,d}$:
	\begin{align}
	C_{2,d}^{(w)}[\eta,\eta';R]=\,&-a[\eta]^{-\frac {d-2}2}\int^\eta_{\eta'} \dd \eta_2\,a[\eta_2]^{d-2}\int^{\eta_2}_{\eta'} \dd \eta_1\,a[\eta_1]^{-\frac {d-2}2}C^{(w)}_{1,d}[\eta_1,\eta';R] \notag\\
	&-w^{\frac{d-3}2}G^{(\mathrm E)}_d[R]\left(\frac{a[\eta']}{a[\eta]}\right)^{\frac {d-2}2}\int^\eta_{\eta'} \dd\eta_1\left(\frac{a[\eta_1]}{a[\eta']}\right)^{d-2},
	\label{C2_Psi_TimeIntegral_Cosmology}
	\end{align}
and the solutions of the retarded scalar Green's function $G^{(w,+)}_d$ for $d\geq 4$, obeying eq.~\eqref{WaveEq_ScalarGreensFunction} with $\kappa = (d - 2)/q_w$ and the re-scaled spatial coordinates: $\vec x \to \vec x / \sqrt w$ and $\vec x' \to \vec x' / \sqrt w$, may be found in eqs.~(131) and (132) of \cite{Chu:2016ngc} -- for even dimensions,
\begin{align}
G^{(w,+)}_{\text{even }d\geq4}[\eta,\eta';R] &  = - \Theta[T] \left(  \frac1{2\pi} \frac{\partial}{\partial \overline \sigma_w} \right)^{\frac{d-2}2} \left( \frac{\Theta[\overline\sigma_w]}2 P_{ \frac{d-2}{q_w} }\left[1+\frac{\overline\sigma_w}{\eta\eta'}\right]\right) , \quad \overline\sigma_w = \frac{(\eta - \eta')^2 - R^2/w}2 .
\label{AcousticGreensFunction_EvenDimensions}
\end{align}
To arrive at eq.~\eqref{BardeenPsi_RadiationDominated} from eq.~(264) of \cite{Chu:2019ndv}, we have carried out the time derivatives within the latter, by making use of the homogeneous wave equation obeyed by $C^{(w)}_{2,d}$ and $C^{(w)}_{3,d}$ to convert their higher derivatives into their lower ones, then taking appropriate linear combinations to simplify the ensuing expressions.

With the solution of $\Psi$ at hand, the other Bardeen scalar $\Phi$ can be obtained immediately by inserting  eq.~\eqref{BardeenPsi_RadiationDominated} into the relation \eqref{PhiPsiRelationship}, or its localized version given in eq.~(175) of \cite{Chu:2019ndv},
\begin{align}
\Phi[\eta,\vec x]=\,&(d-3)\Psi[\eta,\vec x] + \frac{8\pi G_\text N}{d-2} \int_{\mathbb{R}^{d-1}}  \dd^{d-1}\vec{x}'  \left(
G^{\mathrm{(E)}}_{d}{}^{(\text{a})}T_{ll}[\eta,\vec{x}']
-(d-1)\partial_i\partial_jD_d{}^{(\text{a})}T_{ij}[\eta,\vec{x}']\right).
\label{PhiPsiRelationship_Convolution}
\end{align}
{\bf Remarks on Acoustic Modes} \qquad We highlight here, according to eq.~\eqref{Psi_WaveEq_Cosmology}, the Bardeen gauge-invariant scalars obey wave equations -- albeit with associated acoustic cones $0 < |\dd \vec{x}|/\dd\eta = \sqrt{w} \leq 1$ instead of the null one of the spin-2 counterpart in eq.~\eqref{Spin-2_WaveEquation_Cosmology}. This distinct cone structure of the acoustic waves is due to the $\overline{\sigma}_w$ dependence in eq.~\eqref{AcousticGreensFunction_EvenDimensions}, as opposed to the $\overline{\sigma}$ in eq.~\eqref{GeneralSol_1stGreensFunction_EvenDimensions}. It is precisely these acoustic modes that prompted us to examine whether their wave solutions in \eqref{BardeenPsi_RadiationDominated} and \eqref{PhiPsiRelationship_Convolution} are mere artifacts of the decoupling procedure employed in \cite{Chu:2016ngc} to obtain standalone equations for each and every metric variable; or, whether these acoustic features do in fact contribute to the traceless portion of the physical tidal forces in eq.~\eqref{TidalForces}. We now turn to this question, for all even dimensional ($d \geq 4$) relativistic cosmologies.

{\bf Linearized Weyl Tensor} \qquad In terms of the gauge-invariant variables in eqs.~\eqref{Bardeen_Psi} through \eqref{Bardeen_VandDij}, the linearized Weyl components we are after read
\begin{align}
\delta_1 C^i{}_{0j0}=\,&\left(\frac{d-3}{d-2}\right)\Bigg\{\left(\partial_i\partial_j-\frac{\delta_{ij}}{d-1}\vec\nabla^2\right)
\left(\Phi+\Psi\right)+\partial_{(i}\dot{V}_{j)}-\frac12\bigg(\ddot{D}_{ij}+\frac1{d-3}\vec\nabla^2D_{ij}\bigg)\Bigg\} .
\label{LinearizedWeyl_GaugeInvariant}
\end{align}
To compute them, we proceed to substitute the gauge-invariant solutions \eqref{Spin2Solution_RadiationDominated}, \eqref{VectorMode_RadiationDominated}, \eqref{BardeenPsi_RadiationDominated}, and \eqref{PhiPsiRelationship_Convolution} into eq.~\eqref{LinearizedWeyl_GaugeInvariant}. Since these $\delta_1 C^i{}_{0j0}$ are the traceless components of the physical tidal tensor, the reader would not be surprised to learn that eq.~\eqref{LinearizedWeyl_GaugeInvariant} has been shown in \cite{Chu:2019ndv} to be causal -- namely, they depend on the astrophysical system on or inside the null cone of the observer at $(\eta,\vec{x})$. The first attempt made in \cite{Chu:2019ndv} revealed: all the gauge invariant variables $\{ \Phi,\Psi,V_i,D_{ij} \}$ were needed to ensure this causality to hold, because each and every one of them are acausal -- but when inserted into eq.~\eqref{LinearizedWeyl_GaugeInvariant}, the acausal terms of these gauge-invariant variables cancel among themselves.

Moreover, we are able to simplify somewhat the result for $\delta_1 C^i{}_{0j0}$ relative to that in \cite{Chu:2019ndv}. We have already noted so for the $\Psi$ in eq.~\eqref{BardeenPsi_RadiationDominated}. The contributions from $D_{ij}$ can also be simplified by converting the second time derivatives of $C^{(g)}_{2,d}$ and $C^{(g)}_{3,d}$ to their lower-derivative terms using the homogeneous wave equation they satisfy, followed by replacing certain time integrals of $C^{(g)}_{1,d}$ with the new commutators $C^{(g)}_{V,d}$ and $C^{(g)}_{S,d}$. After all these steps, the full $\delta_1C^i{}_{0j0}$ can be expressed in terms of the retarded scalar Green's functions $ G^{(g,+)}_d = -\Theta[T]C^{(g)}_{1,d}$, $ G^{(V,+)}_d = -\Theta[T]C^{(g)}_{V,d}$, and $ G^{(S,+)}_d = -\Theta[T]C^{(g)}_{S,d}$ in the spin-2 sector and $ G^{(w,+)}_d = -\Theta[T]C^{(w)}_{1,d}$ for the Bardeen scalars.
{\allowdisplaybreaks
	\begin{align}
	\delta_1C^i{}_{0j0}[\eta,\vec x]
	&= \delta_1 C^{(g)i}{}_{0j0}[\eta,\vec x] + \delta_1 C^{(\Psi)i}{}_{0j0}[\eta,\vec x] \nonumber\\
	& + \frac{8\pi G_\text N}{d-2} \left({}^{(\text{a})}T_{ij}[\eta,\vec x]-\frac{\delta_{ij}}{d-1}\left((d-3){}^{(\text{a})}T_{00}[\eta,\vec x]
	+2{}^{(\text{a})}T_{ll}[\eta,\vec x]\right)\right) ,
	\label{WeylTensor_RelativisticFluid_General}
	\end{align}}%
where the tensor-only contribution is
{\allowdisplaybreaks\begin{align}
	\delta_1 C^{(g)i}{}_{0j0}&[\eta,\vec x]
	= 8\pi G_\text N   \int_{\mathbb{R}^{d-1}} \dd^{d-1}\vec x' \int_0^\infty  \dd\eta' \,\left(\frac{a[\eta']}{a[\eta]}\right)^{\frac {d-2}2}  \Bigg\{  \bigg(\ddot{ G}^{(g,+)}_{d}-(d-3)\mathcal H[\eta]\dot{ G}^{(g,+)}_{d}    \notag\\
	&+\frac{(d-2)(d-4+ q_w)}4\mathcal H[\eta]^2  G^{(g,+)}_{d}\bigg)  \left({}^{(\text{a})}  T_{ij}[\eta',\vec x']+\frac{\delta_{ij}}{d-2}  \left({}^{(\text{a})}  T_{00}[\eta',\vec x'] - {}^{(\text{a})} T_{ll}[\eta',\vec x']  \right) \right) \notag\\
	& - 2 a[\eta]^{\frac{d-4}2}  \partial_\eta \left( a[\eta]^{-\frac{d-4}2} \partial_{(i}G^{(V,+)}_{d} \right) {}^{(\text{a})} T_{j)0}[\eta',\vec x']   + \frac{\delta_{ij}}{d-2} \mathcal H[\eta']   a[\eta]^{\frac{d-4}2} \partial_\eta \left( a[\eta]^{-\frac{d-4}2} G^{(V,+)}_{d} \right) \notag\\
& \times \left( (d-3) {}^{(\text{a})} T_{00}[\eta',\vec x']  +  {}^{(\text{a})} T_{ll}[\eta',\vec x'] \right)  + \frac{1}{d-2}\Bigg( \frac{d-3}{d-2+\frac{q_w}2} \bigg(  \partial_i\partial_j G^{(S,+)}_{d} + \bigg( \frac{1+\frac{q_w}2}{d-3}\bigg)  \partial_i\partial_j G^{(g,+)}_{d}  \bigg)   \notag\\
	&+ (d-3)\left(1+\frac{q_w}2\right)  \mathcal H[\eta] \mathcal H[\eta'] \partial_i\partial_j   Q^{(V,+)}_{d} \Bigg)  \left( (d-3){}^{(\text{a})}  T_{00}[\eta',\vec x'] + {}^{(\text{a})}  T_{ll}[\eta',\vec x']  \right) \Bigg\} ,
\label{WeylTensor_RelativisticFluid_General_TensorOnly}
	\end{align}}
and that of the scalar-only ones is
\begin{align}
	\delta_1 C^{(\Psi)i}{}_{0j0}[\eta,\vec x]
	&= 8\pi G_\text N  \left(\frac{d-3}{d-2}\right) \left(1+\frac{q_w}2\right) \frac1{w^{\frac{d-1}{2}}}  \int_{\mathbb{R}^{d-1}} \dd^{d-1}\vec x' \int_0^\infty  \dd\eta' \,\left(\frac{a[\eta']}{a[\eta]}\right)^{\frac {d-2}2} \mathcal H[\eta] \mathcal H[\eta'] \nonumber\\
	&\qquad\times \left(
	\frac{\delta_{ij}}{d-1} G^{(w,+)}_{d} - w \partial_i\partial_j  Q^{(w,+)}_{d} \right) \left( (d-3) {}^{(\text{a})} T_{00}[\eta',\vec x']  +  {}^{(\text{a})} T_{ll}[\eta',\vec x'] \right) .
\label{WeylTensor_RelativisticFluid_General_ScalarOnly}
\end{align}
Here, the exact even $d\geq4$ solutions of $G^{(V,+)}_d$ and $ G^{(S,+)}_d$, obeying the wave equation \eqref{WaveEq_ScalarGreensFunction} for $\kappa = (d - 2)/q_w$ and $\kappa = 1 + (d - 2)/q_w$, respectively, can be derived similarly to both $ G^{(g,+)}_d$ and $ G^{(w,+)}_d$ through Nariai's ansatz delineated in appendix (D) of \cite{Chu:2016ngc},
{\allowdisplaybreaks\begin{align}
\label{GeneralSol_2ndGreensFunction_EvenDimensions}
G^{(V,+)}_{\text{even }d\geq4}[\eta,\eta';R] &  = - \Theta[T] \left(  \frac1{2\pi} \frac{\partial}{\partial \overline \sigma} \right)^{\frac{d-2}2} \left( \frac{\Theta[\overline\sigma]}2 P_{  \frac{d-2}{q_w} }\left[1+\frac{\overline\sigma}{\eta\eta'}\right]\right), \qquad
\overline\sigma = \frac{(\eta - \eta')^2 - R^2}2 ; \\
\label{GeneralSol_3rdGreensFunction_EvenDimensions}
G^{(S,+)}_{\text{even }d\geq4}[\eta,\eta';R] &  = - \Theta[T] \left(  \frac1{2\pi} \frac{\partial}{\partial \overline \sigma} \right)^{\frac{d-2}2} \left( \frac{\Theta[\overline\sigma]}2 P_{ 1 + \frac{d-2}{q_w} }\left[1+\frac{\overline\sigma}{\eta\eta'}\right]\right).
\end{align}}%
The remaining time integrals in the effective Green's function of $\delta_1 C^i{}_{0j0}$ are encapsulated within $Q^{(V,+)}_{d}$ and $Q^{(w,+)}_{d}$, defined by
{\allowdisplaybreaks
	\begin{align}
	Q^{(V,+)}_d[\eta,\eta';R] & \equiv  a[\eta]^{-\frac {d-2}2}    \int^\eta_{\eta'} \dd \eta_2\,a[\eta_2]^{d-2}\int^{\eta_2}_{\eta'} \dd \eta_1\,a[\eta_1]^{-\frac {d-2}2}    G^{(V,+)}_{d}[\eta_1,\eta';R] ,
	\label{QV_d}\\
	Q^{(w,+)}_d[\eta,\eta';R] & \equiv  a[\eta]^{-\frac {d-2}2}   \int^\eta_{\eta'} \dd \eta_2\,a[\eta_2]^{d-2}\int^{\eta_2}_{\eta'} \dd \eta_1\,a[\eta_1]^{-\frac {d-2}2}   G^{(w,+)}_{d}[\eta_1,\eta';R].
	\label{Qw_d}
	\end{align}}%
Note that $G^{(w,+)}_d$ and $Q^{(w,+)}_d$ may be obtained from $G^{(V,+)}_d$ and $Q^{(V,+)}_d$ simply via the replacement $R \to R/\sqrt{w}$.\footnote{We have further checked the computation in eq.~\eqref{WeylTensor_RelativisticFluid_General} by performing it in Fourier ($\vec{k}$-)space.}

At this juncture, it is the time integrals in eqs.~\eqref{QV_d} and \eqref{Qw_d} that are currently the primary obstacles towards an explicit closed arbitrary$-w$ expression for $\delta_1C^i{}_{0j0}$ in eq.~\eqref{WeylTensor_RelativisticFluid_General}. Nonetheless, these integrals may be evaluated for particular equation-of-states. The following section, i.e., \S \eqref{PolarizationPattern_4DRadiation} below, will focus exclusively on the radiation dominated $w=1/3$ case. The de Sitter $w=-1$ and matter dominated $w=0$ cases may be found in appendix \eqref{LinearizedWeyl_deSitterAndMatterDomination}.

The physical significance of the result in eq.~\eqref{WeylTensor_RelativisticFluid_General} is the appearance of the acoustic-cone structure encoded in $G^{(w,+)}_{d}$ and $Q^{(w,+)}_{d}$. That is, the acoustic waves of $\Phi$ and $\Psi$ do not appear to be mere artifacts of the decoupling procedure of \cite{Chu:2016ngc}; rather, eq.~\eqref{WeylTensor_RelativisticFluid_General} instead tells us the trace-free tidal forces in a relativistic cosmology $0 < w \leq 1$ do in fact carry acoustic modes that propagate at equal to or less than speed $\sqrt{w}$. Of course, in the limit of a very dilute universe, we expect to recover Minkowski spacetime -- these acoustic tidal forces must therefore be Hubble-suppressed relative to the tensor ones.

More explicitly, provided that the observer is located away from the GW sources ($R\neq 0$), the acoustic-only tidal forces \eqref{WeylTensor_RelativisticFluid_General_ScalarOnly} can be re-cast into
{\allowdisplaybreaks
	\begin{align}
	\delta_1 C^{(\Psi)i}{}_{0j0}[\eta,\vec x] & =  8\pi G_\text N  \left(\frac{d-3}{d-2}\right)\left(1+\frac{q_w}2\right) \frac1{w^{\frac{d-1}2}}  \int_{\mathbb{R}^{d-1}} \dd^{d-1}\vec x' \int_0^\infty  \dd\eta' \,\left(\frac{a[\eta']}{a[\eta]}\right)^{\frac {d-2}{2}} \left( \delta_{ij} - (d-1) \widehat R_i\widehat R_j \right) \notag\\
	&\times\mathcal H[\eta] \mathcal H[\eta']  \bigg( \frac1{d-1} \, G^{(w,+)}_{d} - \frac wR\frac{\partial}{\partial R} Q^{(w,+)}_d \bigg)\left( (d-3){}^{(\text{a})}  T_{00}[\eta',\vec x'] + {}^{(\text{a})}  T_{ll}[\eta',\vec x'] \right) .
	\label{Weyl_ScalarContribution}
	\end{align}}%
Notice the $\delta_1 C^{(\Psi)i}{}_{0j0}$ in eq.~\eqref{Weyl_ScalarContribution} is manifestly trace-less due to the $\delta_{ij} - (d-1) \widehat R_i\widehat R_j$; this is consistent with the Weyl being the trace-free part of Riemann. To arrive at eq.~\eqref{Weyl_ScalarContribution}, we have also employed the field equation obeyed by $C^{(w)}_{1,d}$ to convert $\partial_i\partial_jQ^{(w,+)}_d$ into
\begin{align}
\partial_i\partial_j  Q^{(w,+)}_d =\,&  \widehat R_i\widehat R_j w^{-1}  G^{(w,+)}_{d}  + \left( \delta_{ij} - (d-1) \widehat R_i\widehat R_j \right) \frac1R\frac{\partial}{\partial R} Q^{(w,+)}_d , \qquad  (R\neq0).
\label{DoubleSpatialDerivatives_Qw}
\end{align}
That the Green's functions of $\Psi$ in eq.~\eqref{BardeenPsi_RadiationDominated} and $\Phi$ in eq.~\eqref{PhiPsiRelationship_Convolution} depend on $C_{2,d}^{(w)}$ but not on $C_{1,d}^{(w)}$ helps ensure the acoustic contributions to the Weyl tensor in eq.~\eqref{Weyl_ScalarContribution} are Hubble-suppressed relative to the tensor counterparts -- i.e., the $\mathcal H[\eta] \mathcal H[\eta']$ factors tell us, as already alluded to above, as the universe dilutes ($\mathcal{H} \to 0$) we should recover the Minkowski limit, where these acoustic waves should cease to exist. Furthermore, we will witness in some detail below, the acoustic-cone scalar signal to the null-cone tensor signal scales as $(H \tau_c)^2 \ll 1$, where $H \equiv \dot a/a^2 = \mathcal H / a$ is the usual Hubble parameter and $\tau_c$ the characteristic timescale of the source.

Let us now move on in the following section to examine the features of these acoustic-gravitational waves for the physically important radiation dominated phase of our 4D universe.

\section{Gravitational Waves and Polarization Patterns in 4D Radiation-Dominated Universe} \label{PolarizationPattern_4DRadiation}

When specialized to 4D radiation domination
\begin{align}
a[\eta] = \frac{\eta}{\eta_0}, \qquad d=4, \ w = \frac{1}{3} ;
\end{align}
the full exact $\delta_1 C^i{}_{0j0}$ in eq.~\eqref{WeylTensor_RelativisticFluid_General} reads
	\begin{align}
	\delta_1C^{(\text{4D rad})i}{}_{0j0}[\eta,\vec x]
	&= \delta_1 C^{(\text{$g \vert $4D rad})i}{}_{0j0}[\eta,\vec x] + \delta_1 C^{(\text{$\Psi|$4D rad})i}{}_{0j0}[\eta,\vec x] \nonumber\\
	& + 4\pi G_\text N \left({}^{(\text{a})}T_{ij}[\eta,\vec x] - \frac{\delta_{ij}}{3}\left({}^{(\text{a})}T_{00}[\eta,\vec x]
	+2{}^{(\text{a})}T_{ll}[\eta, \vec x]\right)\right) ;
	\label{WeylTensor_4DRadiationDomination}
	\end{align}
where the tensor-only portions are
{\allowdisplaybreaks
\begin{align}
\delta_1  C^{(\text{$g \vert $4D rad})i}&{}_{0j0} [\eta,\vec x]  =  8\pi G_\text N  \int_{\mathbb{R}^{3}} \dd^{3}\vec x' \int_0^\infty  \dd\eta' \,\left(\frac{\eta'}{\eta}\right) \Bigg\{  \bigg(\ddot{ G}^{(g|\text{rad},+)}_{4}-\frac1\eta\dot{ G}^{(g|\text{rad},+)}_{4}    + \frac1{\eta^2}  G^{(g|\text{rad},+)}_{4}\bigg)  \notag\\
& \times \left({}^{(\text{a})}  T_{ij}[\eta',\vec x']+\frac{\delta_{ij}}{2}  \left({}^{(\text{a})}  T_{00}[\eta',\vec x'] - {}^{(\text{a})} T_{ll}[\eta',\vec x']  \right) \right) - 2     \partial_{(i}\dot G^{(V|\text{rad},+)}_{4}  {}^{(\text{a})} T_{j)0}[\eta',\vec x'] \notag\\
& + \frac{\delta_{ij}}{2\eta'}    \dot  G^{(V|\text{rad},+)}_{4} \left(  {}^{(\text{a})} T_{00}[\eta',\vec x']  +  {}^{(\text{a})} T_{ll}[\eta',\vec x'] \right)  + \bigg( \frac{1}{6}  \left(  \partial_i\partial_j G^{(S|\text{rad},+)}_{4} + 2  \partial_i\partial_j G^{(g|\text{rad},+)}_{4}  \right) \notag\\
&  + \frac1{\eta\eta'}  \partial_i\partial_j Q^{(V|\text{rad},+)}_4  \bigg)  \left( {}^{(\text{a})}  T_{00}[\eta',\vec x'] + {}^{(\text{a})}  T_{ll}[\eta',\vec x']  \right) \Bigg\} ,
\label{WeylTensor_4DRadiationDomination_TensorOnly}
\end{align}}%
and the scalar-only portions are
{\allowdisplaybreaks
	\begin{align}
	\delta_1  C^{(\text{$\Psi|$4D rad})i}&{}_{0j0} [\eta,\vec x]  =  8\pi G_\text N  \int_{\mathbb{R}^{3}} \dd^{3}\vec x' \int_0^\infty  \dd\eta' \, \frac{\sqrt{3}}{\eta^2} \notag \\
	&\qquad\times \left(
	\delta_{ij} G^{(w|\text{rad},+)}_{4} - \partial_i\partial_j  Q^{(w|\text{rad},+)}_4 \right) \left( {}^{(\text{a})}  T_{00}[\eta',\vec x'] + {}^{(\text{a})}  T_{ll}[\eta',\vec x']  \right) .
\label{WeylTensor_4DRadiationDomination_ScalarOnly}
	\end{align}}%
The relevant retarded scalar Green's functions $G^{(g|\text{rad},+)}_{4}$, $G^{(w|\text{rad},+)}_{4}$, $G^{(V|\text{rad},+)}_{4}$, and $G^{(S|\text{rad},+)}_{4}$, respectively, are given in eqs.~\eqref{GeneralSol_1stGreensFunction_EvenDimensions}, \eqref{AcousticGreensFunction_EvenDimensions}, \eqref{GeneralSol_2ndGreensFunction_EvenDimensions}, and \eqref{GeneralSol_3rdGreensFunction_EvenDimensions} with $d$ set to 4 and $w$ to $1/3$,
{\allowdisplaybreaks
\begin{align}
G^{(g|\text{rad},+)}_{4}[\eta,\eta';R] & = -\frac{\delta[T-R]}{4\pi R} ,
\label{1stScalarGreensFunction_Spin2_4DRadiation}\\
G^{(V|\text{rad},+)}_{4}[\eta,\eta';R] & = -\frac{\delta[T-R]}{4\pi R} - \frac{\Theta[T-R]}{4\pi \eta\eta'}  ,
\label{2ndScalarGreensFunction_Spin2_4DRadiation} \\
G^{(S|\text{rad},+)}_{4}[\eta,\eta';R] & = -\frac{\delta[T-R]}{4\pi R}  - \frac{3 \Theta[T-R]}{4\pi\eta\eta'} \left(1 + \frac{(\eta - \eta')^2 - R^2}{2\eta\eta'} \right)  ,
\label{3rdScalarGreensFunction_Spin2_4DRadiation} \\
G^{(w|\text{rad},+)}_{4}[\eta,\eta';R] & =  -\frac{ \delta \big[T-\sqrt 3 R \big]}{4 \sqrt 3 \, \pi R} - \frac{\Theta\big[T-\sqrt 3 R \big]}{4\pi \eta\eta'} .
\label{1stScalarGreensFunction_Bardeen_4DRadiation}
\end{align}}%
The $Q^{(V|\text{rad},+)}_{4}$ and $Q^{(w|\text{rad},+)}_{4}$ from eqs.~\eqref{QV_d} and \eqref{Qw_d} can now be analytically evaluated:
{\allowdisplaybreaks
\begin{align}
Q^{(V|\text{rad},+)}_{4}[\eta,\eta';R] & =  - \frac{\Theta[T-R]}{24\pi \eta \eta' } \bigg( R^2 - 3\left(\eta^2 + \eta'^2\right)  + \frac{2\big(\eta^3 - \eta'^3\big)}{R}\bigg) ,
\label{QV_4DRadiation} \\
Q^{(w|\text{rad},+)}_{4}[\eta,\eta';R] & =   - \frac{\Theta\big[T-\sqrt 3 R\big]}{24\pi \eta \eta'} \bigg( 3R^2 - 3\left(\eta^2 + \eta'^2\right)  + \frac{2\big(\eta^3 - \eta'^3\big)}{\sqrt 3 R}\bigg) .
\label{Qw_4DRadiation}
\end{align}}%
They are both pure-tail and vanish identically on the null/acoustic cones. We notice that the pure light-cone nature of $G^{(g|\text{rad},+)}_{4}$ in eq.~\eqref{1stScalarGreensFunction_Spin2_4DRadiation} is closely tied to its conformal invariance. Whereas the other scalar Green's functions in eqs.~\eqref{2ndScalarGreensFunction_Spin2_4DRadiation}, \eqref{3rdScalarGreensFunction_Spin2_4DRadiation}, and \eqref{1stScalarGreensFunction_Bardeen_4DRadiation} all develop tails. Additionally, the tail functions of $G^{(V|\text{rad},+)}_{4}$ and $G^{(w|\text{rad},+)}_{4}$ are both space-independent.

Assuming the observer at $(\eta,\vec x)$ is away from the source area ($R\neq 0$), the scalar-only contributions in eq.~\eqref{WeylTensor_4DRadiationDomination_ScalarOnly} are now, according to eq.~\eqref{Weyl_ScalarContribution},
\begin{align}
\delta_1 C^{(\Psi|\text{4D rad})i}{}_{0j0}[\eta,\vec x] & = -2 G_\text N \int_{\mathbb{R}^{3}} \dd^{3} \vec x' \left(\delta_{ij}-3\widehat R_i\widehat R_j\right)  \Bigg\{ \frac1{R\eta^2} \left( {}^{(\text{a})}  T_{00}\big[\eta - \sqrt 3 R,\vec x'\big] + {}^{(\text{a})}  T_{ll} \big[\eta - \sqrt 3 R,\vec x' \big]  \right)  \notag\\
& + \frac1{3R^3}\int_0^{\eta - \sqrt 3 R - 0^+}  \dd\eta' \, \frac{(\eta^3 - \eta'^3)}{ \eta^3 \eta'}  \left( {}^{(\text{a})}  T_{00}[\eta',\vec x'] + {}^{(\text{a})}  T_{ll}[\eta',\vec x']  \right)  \Bigg\} .
\label{Weyl_ScalarContribution_4DRadiation}
\end{align}
The first line, with the $1/R$ scaling, is composed of the acoustic-cone signals; while the second line, decaying as $1/R^3$ instead, corresponds to the acoustic tails that depend on the entire past history of the source(s) right up to the retarded time $ \eta - \sqrt 3 R$. Even though the tail signals, for a fixed observer time $\eta$, fall off faster with increasing distance than their acoustic-cone counterpart, the former in the far zone is not always guaranteed to be strongly suppressed relative to the latter -- we will carry out the relevant estimates below.

\subsection{Far-Zone JWKB Limit: Trace-Free Tidal Forces}
\label{JWKB_Tidal}

Since radiation corresponds to the transport of energy-momentum away from its emitter, its study usually takes place in the region of space far from the source; where the characteristic timescale of the source $\tau_c$ -- as well as its typical size $r_c$ -- is much shorter than the proper observer-source spatial distance $a[\eta] r$: namely, $\tau_c \ll a[\eta]r$ and $r_c \ll a[\eta] r$. Additionally, about a cosmological background, this necessarily implies $\tau_c$ is much shorter than that associated with the universe itself $1/H$; since the latter is always the longest time/length scale. This far zone `JKWB limit' is what we wish to consider in this section.

{\bf Estimates} \qquad To extract the leading far-zone contributions of eq.~\eqref{WeylTensor_4DRadiationDomination}, we first discard the source terms evaluated at the observer's spacetime location, i.e., its last line, and carry out the derivatives acting on the relevant scalar Green's functions; whereby certain tail terms will be converted into the null-cone/acoustic-cone pieces as the result of differentiating the step functions.

Let us first compare the various ``direct" spin-2 terms; i.e., its light cone signals. One will find, it is the second derivatives of the delta function $\delta''$, which occurs only in the tensor sector, that yield the dominant signals in the far zone. The reasons are as follows. For the tensor sector, each derivative acting on the $\delta$-functions may be integrated by parts; and the dominant piece of the results are time derivatives $\partial_{\eta'}$ acting on the source ${}^{(\text{a})}  T_{\mu \nu}$. These $ \partial_{\eta'}^n {}^{(\text{a})}  T_{\mu \nu}$ terms scale as $a[\eta']^n {}^{(\text{a})}  T_{\mu \nu}/\tau_c^n$ for $n = 1$ and $n = 2$. A direct calculation then reveals, all the sub-dominant ``direct" signals, in the far zone $R \approx r$, are suppressed by factors of order $H[\eta_r] \tau_{c}$ and $\tau_c/(a[\eta_r]r)$ relative to the leading portion of the $\delta''$ terms, where $\tau_c$ is evaluated at the retarded time $\eta_r \equiv \eta - r$. The $H[\eta_r] \tau_c$ corresponds to the ratio (characteristic timescale of the source)/(cosmic age), which we will assume to be small. There are also factors that scale as $(H[\eta_r] \tau_c) H[\eta] ( a[\eta] r)$; but since the $H[\eta] ( a[\eta] r) \sim $ (physical observer-source distance)/(size of the observable universe) $\ll 1$, we will ignore them. Furthermore, there are `finite size' effects, analogous to the multipole expansion in flat spacetime, that scale as $H[\eta_r] r_c$ and $r_c/(a[\eta_r]r)$.

Next, let us move on to compare the ``direct" scalar signals -- i.e., its acoustic-cone portion -- to its spin-2 counterparts. Denote the dominant light-cone terms in eq.~\eqref{WeylTensor_4DRadiationDomination_TensorOnly} by $\delta_1 C^{(g|\text{DLC})}$:
\begin{align}
\delta_1 C^{(g|\text{DLC})}[\eta,\vec x] & \equiv  G_{\text N} \int_{\mathbb{R}^{3}} \dd^{3}\vec x' \int_0^\infty  \dd\eta' \,\left( \frac{\eta'}\eta \right) \frac{\delta''[T-R]}R S[\eta',\vec x'] \notag\\
& \sim  G_{\text N}  \int_{\mathbb{R}^{3}} \dd^{3}\vec x' \, \left(\frac{\eta - R}\eta\right) \frac{ a[\eta - R]^2 }{\tau_c^2 R } S[\eta - R,\vec x']  ,
\label{DominantLightConeAmplitude_4DRadiation}
\end{align}
where the function $S$ refers to a generic component of the matter stress-energy tensor ${}^{(\text{a})}  T_{\mu \nu}$\footnote{Strictly speaking, the amplitude discrepancies between different components of ${}^{(\text{a})}  T_{\widehat\mu \widehat\nu}$ may emerge in the non-relativistic limit, where the stress components ${}^{(\text{a})}  T_{\widehat i \widehat j}$, as well as the momentum density ${}^{(\text{a})}  T_{\widehat 0 \widehat i}$,
could be suppressed relative to the energy density ${}^{(\text{a})}  T_{\widehat 0 \widehat 0}$. Note that ${}^{(\text{a})}  T_{\widehat\mu \widehat\nu} \equiv a^{-2} {}^{(\text{a})}  T_{\mu\nu} $ denotes the physical matter stress tensor observed in a co-moving orthonormal frame. However, we will not take into consideration the distinction between the relativistic and non-relativistic cases and only assume they are of the same order in our estimates here.
}, and all the numerical constants and the tensor structure have been omitted. In a similar manner, the acoustic-cone amplitude of eq.~\eqref{Weyl_ScalarContribution_4DRadiation} can also be written schematically as
\begin{align}
\delta_1 C^{(\Psi|\text{direct})}[\eta,\vec x] \equiv G_{\text N}  \int_{\mathbb{R}^{3}} \dd^{3}\vec x' \, \frac1{R\eta^2}  S\big[\eta - \sqrt 3 R,\vec x'\big] .
\label{AcousticConeAmlitude_4DRadiation}
\end{align}
Ignoring the different propagation cones between the tensor versus acoustic signals in eqs.~\eqref{DominantLightConeAmplitude_4DRadiation} and \eqref{AcousticConeAmlitude_4DRadiation}, we take the ratio of their far-zone amplitudes to deduce
\begin{align}
\left|\frac{\delta_1 C^{(\Psi|\text{direct})}[\eta,\vec x]}{\delta_1 C^{(g|\text{DLC})}[\eta,\vec x]}\right| \sim  \big( H[\eta_r] \tau_c\big)^2 , \qquad (\text{far zone}) .
\label{AcousticToNull_Direct_4DRadiation}
\end{align}
In addition to the ``direct'' part of the signal, there also exist non-vanishing tail effects in $\delta_1  C^{(\text{4D rad})i}{}_{0j0} $, as opposed to the de Sitter case (see appendix \eqref{LinearizedWeyl_deSitterAndMatterDomination}). More specifically, the spin-2 tail portion of eq.~\eqref{WeylTensor_4DRadiationDomination_TensorOnly} generally contains two types of amplitudes:
{\allowdisplaybreaks
\begin{align}
\delta_1 C^{(g|\text{tail-1})}[\eta,\vec x] &\equiv G_\text N \int_{\mathbb{R}^{3}} \dd^{3}\vec x' \int_0^{\eta - R - 0^+}  \dd\eta' \, \frac1{\eta^3 \eta'} S[\eta',\vec x'] ,
\label{Spin2TailAmlitude1_4DRadiation}\\
\delta_1 C^{(g|\text{tail-2})}[\eta,\vec x] &\equiv  G_\text N \int_{\mathbb{R}^{3}} \dd^{3}\vec x' \frac1{R^3}\int_0^{\eta - R - 0^+} \dd\eta' \, \frac{(\eta^3 - \eta'^3)}{\eta^3 \eta'} S[\eta',\vec x'] ,
\label{Spin2TailAmlitude2_4DRadiation}
\end{align}}%
while the acoustic tail, as we have already noted in eq.~\eqref{Weyl_ScalarContribution_4DRadiation}, has the same amplitude as the latter up to a re-scaled spatial dependence $R \to \sqrt 3 R$,
\begin{align}
\delta_1 C^{(\Psi|\text{tail})}[\eta,\vec x] &\equiv  G_\text N \int_{\mathbb{R}^{3}} \dd^{3}\vec x' \frac1{R^3}\int_0^{\eta - \sqrt 3 R - 0^+} \dd\eta' \, \frac{(\eta^3 - \eta'^3)}{\eta^3 \eta'} S[\eta',\vec x'] .
\label{AcousticTailAmlitude_4DRadiation}
\end{align}
Within these tail terms, the $S[\eta',\vec x']$ usually involves both ${}^{(\text{a})}  T_{00}[\eta',\vec x']$ and ${}^{(\text{a})}  T_{ll}[\eta',\vec x']$ of the matter source; however, the former, in the far-zone limit, may potentially lead to divergent tail integrals in eqs.~\eqref{Spin2TailAmlitude1_4DRadiation}, \eqref{Spin2TailAmlitude2_4DRadiation}, and \eqref{AcousticTailAmlitude_4DRadiation}. This is because, the total mass/energy of the astrophysical system ${}^{(\text{a})}M$,
\begin{align}
{}^{(\text{a})}M[\eta] & \equiv   \int_{\mathbb{R}^{3}} \dd^{3} \vec x \, a[\eta]^3  {}^{(\text{a})}  T_{\widehat 0 \widehat 0}[\eta,\vec x] =   \int_{\mathbb{R}^{3}} \dd^{3} \vec x \, a[\eta]  {}^{(\text{a})}  T_{00}[\eta,\vec x],
\label{TotalMass_Cosmology}
\end{align}
is approximately conserved, at least for small Hubble scales, so that the integral
\begin{align}
\int_{\mathbb{R}^{3}} \dd^{3}\vec x' \int_0^{\eta - R - 0^+}  \dd\eta' \, \frac1{ \eta'} {}^{(\text{a})}  T_{00}[\eta',\vec x'] \sim  {}^{(\text{a})}M  \int_0^{\eta - r - 0^+}  \dd\eta' \, \frac{\eta_0}{\eta'^2} \qquad (\text{far zone}),
\end{align}
as well as its acoustic counterpart in eq.~\eqref{AcousticTailAmlitude_4DRadiation}, appears to blow up at the lower limit $\eta' \to 0$. The notation ${}^{(\text{a})}  T_{\widehat0 \widehat0} \equiv a^{-2} {}^{(\text{a})}  T_{00} $ here denotes the energy density observed by a co-moving observer in a orthonormal frame. As an example, this pathology occurs for co-moving point particles. Whether this divergence poses a real physical issue, however, cannot be clarified until the explicit ${}^{(\text{a})}  T_{\mu\nu}$ of the astrophysical source(s) is specified, which we shall leave to our future work.

In spite of this potential issue, we may estimate the far-zone tail-to-cone amplitude by comparing the signals received solely over the course of the GW generation process -- a physical estimation scheme employed in \cite{Chu:2015yua}. Since the gravitational radiation is physically attributed to the pressure of the astrophysical system due to the work done within itself, the GW signals are expected to peak at some time $\eta_*$ in the past, within a finite duration $\Delta t \sim \int_{\text{peak width}}\dd \eta \, a[\eta]$ of the active GW production from the source. Hence, as far as the radiation process is concerned, with the assumption that the scale factor does not change appreciably over the active period $\Delta t$, the dominant spin-2 light-cone amplitude  \eqref{DominantLightConeAmplitude_4DRadiation} is roughly bounded by
\begin{align}
\left|\delta_1 C^{(g|\text{DLC})}[\eta,\vec x] \right| \lesssim  \frac{G_{\text N}}{\tau_{c*}^2 r } \cdot \frac{   \eta_*^5}{\eta \eta_0^4}  \left| \int_{\mathbb{R}^{3}} \dd^{3}\vec x' \,  \widehat S[\eta_*,\vec x'] \right| ,
\label{DominantLightConeAmplitude_4DRadiation_UpperBound}
\end{align}
whereas the tail ones in eqs.~\eqref{Spin2TailAmlitude1_4DRadiation}, \eqref{Spin2TailAmlitude2_4DRadiation} and \eqref{AcousticTailAmlitude_4DRadiation} are, respectively, bounded by
{\allowdisplaybreaks
\begin{align}
\left|\delta_1 C^{(g|\text{tail-1})}[\eta,\vec x] \right| &  \lesssim  G_\text N \Delta t \cdot \frac{1}{\eta^3 \eta_0}  \left| \int_{\mathbb{R}^{3}} \dd^{3}\vec x' \, \widehat S[\eta_*,\vec x'] \right| ,
\label{Spin2TailAmlitude1_4DRadiation_UpperBound}\\
\left| \delta_1 C^{(g|\text{tail-2})}[\eta,\vec x] \right|   \sim \left| \delta_1 C^{(\Psi|\text{tail})}[\eta,\vec x] \right| & \lesssim  \frac{G_\text N \Delta t}{r^3} \cdot \frac{(\eta^3 - \eta_*^3)}{\eta^3 \eta_0}   \left| \int_{\mathbb{R}^{3}} \dd^{3}\vec x' \,  \widehat S[\eta_*,\vec x']  \right| ,
\label{Spin2TailAmlitude2_4DRadiation_UpperBound}
\end{align}}%
where the far-zone limit has been taken, $\tau_{c*}$ denotes the timescales of the source at the peak time $\eta_* $, and $\widehat S = a^{-2} S$ refers symbolically to the physical ${}^{(\text{a})}  T_{\widehat\mu \widehat\nu} = a^{-2} {}^{(\text{a})}  T_{\mu\nu} $ observed by a co-moving observer. As it turns out, the tail bounds \eqref{Spin2TailAmlitude1_4DRadiation_UpperBound} and \eqref{Spin2TailAmlitude2_4DRadiation_UpperBound}, in the far-zone regime, are highly suppressed relative to their leading null-cone counterpart \eqref{DominantLightConeAmplitude_4DRadiation_UpperBound} by the following ratios,
{\allowdisplaybreaks
\begin{align}
\left|\frac{\delta_1 C^{(g|\text{tail-1})}[\eta,\vec x]}{\delta_1 C^{(g|\text{DLC})}[\eta,\vec x] } \right| & \sim   \big( H[\eta_*]\tau_{c*} \big)^2  \cdot H[\eta_*] \Delta t \cdot H[\eta] \big( a[\eta] r \big) \cdot \frac{a[\eta_*]}{a[\eta]} ,
\label{Spin2TailToCone1_4DRadiation}\\
\left|\frac{\delta_1 C^{(g|\text{tail-2})}[\eta,\vec x]}{\delta_1 C^{(g|\text{DLC})}[\eta,\vec x] } \right| \sim \left|\frac{\delta_1 C^{(\Psi|\text{tail})}[\eta,\vec x]}{\delta_1 C^{(g|\text{DLC})}[\eta,\vec x] } \right| & \sim \left( \frac{\tau_{c*}}{a[\eta_*]r} \right)^2 \cdot H[\eta_*] \Delta t \cdot \frac{a[\eta]}{a[\eta_*]}\left( 1 - \left( \frac{a[\eta_*]}{a[\eta]}\right)^3 \right) .
\label{Spin2TailToCone2_4DRadiation}
\end{align}}%
Note also that the additional suppression factor $H[\eta_*] \Delta t$ scales as (duration of active source)/(cosmic age) at around the peak time $\eta_*$. Strictly speaking, the ratio \eqref{Spin2TailToCone2_4DRadiation} could still be enhanced if the active period of the source took place in the extremely early universe.

Now, having neglected the sub-dominant direct and tail terms in eq.~\eqref{WeylTensor_4DRadiationDomination}, the asymptotic far-zone behavior of $\delta_1  C^{(\text{4D rad})i}{}_{0j0}$ can be further re-cast into a local-in-space ``transverse-traceless'' (``tt'') form, upon integrations by parts and invoking the conservation law of the energy-momentum tensor ($\overline \nabla^\mu \,^{(\text{a})}  T_{\mu\nu} = 0$) at leading order,
{\allowdisplaybreaks
\begin{align}
\partial_i{}^{(\text{a})}T_{ij} = \, &   {}^{(\text{a})} \dot T_{0j}\big( 1+ \mathcal O[H\tau_c] \big) ,
\label{Coserve1_Leading}\\
\partial_j {}^{(\text{a})}T_{0j} = \,&   {}^{(\text{a})}\dot T_{00}\big( 1+ \mathcal O[H\tau_c] \big)  .
\label{Coserve2_Leading}
\end{align}}%
That is, taking into account the scaling estimates performed above, and at leading order, we find that the far-zone JWKB limit of $\delta_1  C^{(\text{4D rad})i}{}_{0j0}$ is simply proportional to the acceleration of the transverse-traceless ``tt'' spatial metric perturbation $\chi_{ij}$ defined in eq.~\eqref{ttGraviton_4DRadiation} below. By placing $\vec{x}=\vec{0}$ within the source, so that $R \approx r \equiv|\vec x|$ in the far zone:
\begin{align}
\label{WeylTensor_4DRadiationDomination_FarZone}
\delta_1  C&^{(\text{4D rad})i}{}_{0j0} [\eta,\vec x] = -\frac12 \ddot \chi^{(\text{tt}|\text{4D rad})}_{ij}[\eta,\vec x] \\
&\qquad\times \Bigg( 1+ \mathcal O\Bigg[
\frac{\tau_c}{ a[\eta_r] r} , \,\,  H[\eta_r] \tau_c , \,\, \frac{r_c}{a[\eta_r]r} , \,\,  H[\eta_r] r_c , \,\,
\left( \frac{\tau_{c*}}{a[\eta_*]r} \right)^2 \cdot H[\eta_*] \Delta t \cdot \frac{a[\eta]}{a[\eta_*]}\left( 1 - \left( \frac{a[\eta_*]}{a[\eta]}\right)^3 \right) \Bigg] \Bigg), \nonumber
\end{align}
where the size $r_c$ and the timescale $\tau_c$ of the source are, strictly speaking, evaluated at the retarded time $\eta_r$, and $\chi^{(\text{tt}|\text{4D rad})}_{ij}$ is given by
\begin{align}
\chi^{(\text{tt}|\text{4D rad})}_{ij}[\eta,\vec x] \equiv P^{(\text{4D})}_{ijmn} \frac{4 G_\text N}r \left(\frac{\eta-r}{\eta}\right) \int_{\mathbb{R}^{3}} \dd^{3}\vec x'  \, {}^{(\text{a})}T_{mn}\big[\eta - r + \vec x' \cdot \widehat r,\vec x'\big],
\label{ttGraviton_4DRadiation}
\end{align}
with $\widehat{r} \equiv \vec{x}/|\vec{x}|$ and $P^{(\text{4D})}_{ijmn}$ being the 4D ``tt'' projection tensor defined in the position space,
{\allowdisplaybreaks
\begin{align}
P^{(\text{4D})}_{ij mn} 	& \equiv P_{m(i} P_{j)n} - \frac{1}{2} P_{ij} P_{mn}, \qquad \quad  P_{ij} \equiv \delta_{ij} - \widehat{r}_i \widehat{r}_j  .
\label{ttProjector}
\end{align}}%
Because of this ``tt'' projector, $\chi^{(\text{tt}|\text{4D rad})}_{ij}$ obeys the conditions $\delta^{ij}\chi^{(\text{tt}|\text{4D rad})}_{ij} = 0 = \widehat r^i\chi^{(\text{tt}|\text{4D rad})}_{ij}$, where the traceless-ness is consistent with that of the Weyl components $\delta_1 C^i{}_{0j0}$, and the transversality implies, at leading order (see eq.~\eqref{WeylTensor_4DRadiationDomination_FarZone}), $\widehat r_i \,\delta_1 C^i{}_{0j0} = 0 = \widehat r^j \,\delta_1 C^i{}_{0j0} $. At this order, we reiterate that the GW tidal forces described in eq.~\eqref{WeylTensor_4DRadiationDomination_FarZone} are exclusively dependent on the spin-2 gravitons, to which the acoustic contributions are highly suppressed in comparison. Moreover, the dominant far-zone behavior of $\delta_1  C^{(\text{4D rad})i}{}_{0j0} $, except the extra redshift factor $(\eta-r)/\eta$ in eq.~\eqref{ttGraviton_4DRadiation}, is closely analogous to its flat-spacetime counterpart. (See, e.g., eqs.~(52) and (199) of \cite{Chu:2019ndv}, and recall in eq.~\eqref{RiemannWeyl} that $C^\mu{}_{\nu\rho\sigma}$ and $R^\mu{}_{\nu\rho\sigma}$, in Minkowski background, are equivalent in a source-free region.) Thus, one would expect the far-zone spin-2 polarization pattern, determined by eq.~\eqref{ttProjector}, to look the same as the Minkowski predictions, which will be further elaborated below. Finally, it is worth highlighting that, the limit taken in eq.~\eqref{WeylTensor_4DRadiationDomination_FarZone} to extract the dominant tidal forces, in essence, coincides with the high-frequency regime of GWs in light of the JWKB approximation. In particular, all the wave tails, encoded in eqs.~\eqref{2ndScalarGreensFunction_Spin2_4DRadiation}, \eqref{3rdScalarGreensFunction_Spin2_4DRadiation}, \eqref{1stScalarGreensFunction_Bardeen_4DRadiation}, \eqref{QV_4DRadiation}, and \eqref{Qw_4DRadiation}, become irrelevant within such a limit.

If the astrophysical system is non-relativistic, $r_c/\tau_c \to 0$, the ``tt'' perturbations in eq.~\eqref{ttGraviton_4DRadiation} may also be re-expressed in terms of the quadrupole moment ${}^{(\text{a})}I_{mn}$, by virtue of the energy-momentum conservation given in eqs.~\eqref{Coserve1_Leading} and \eqref{Coserve2_Leading},
\begin{align}
\chi^{(\text{tt}|\text{4D rad})}_{ij}[\eta,\vec x] \approx P^{(\text{4D})}_{ijmn} \frac{2 G_\text N}{a[\eta]r}  \frac{ 1}{a[\eta-r]^2}  {}^{(\text{a})} \ddot I_{mn}[\eta - r] \qquad (\text{non-relativistic}),
\label{ttGraviton_4DRadiation_Quadrupole}
\end{align}
where the mass quadrupole moment ${}^{(\text{a})}I_{mn}$ is defined as
\begin{align}
{}^{(\text{a})}I_{mn}[\eta] & \equiv \int_{\mathbb{R}^{3}} \dd^{3} \vec x  \, a[\eta]^{3} \left(a[\eta] x^i\right) \left( a[\eta] x^j \right) {}^{(\text{a})}  T_{\widehat 0\widehat 0}[\eta,\vec x] = \int_{\mathbb{R}^{3}} \dd^{3} \vec x  \, a[\eta]^{3}  x^i  x^j {}^{(\text{a})}  T_{00}[\eta,\vec x].
\label{QuadrupoleMoment_Cosmology}
\end{align}

{\bf Scalar Acoustic-Gravitational Tidal Forces} \qquad Although the Bardeen scalars contribute sub-dominantly to the far-zone $\delta_1  C^{(\text{4D rad})i}{}_{0j0}$, their mere presence does raise the question of how many dynamical degrees-of-freedom there are within the linearized Einstein-fluid equations at hand. The non-trivial acoustic tidal forces also point to their potential impact on the large scale structure of the universe. To this end, we may continue to extract the explicit far-zone behavior of their contribution \eqref{Weyl_ScalarContribution_4DRadiation},
\begin{align}
\delta_1 C^{(\Psi|\text{4D rad})i}{}_{0j0}[\eta,\vec x] =   - \frac12 \Bigg\{ \mathcal S_{ij}^{(\Psi|\text{direct})}[\eta,\vec x] \bigg(  1 & + \mathcal O \left[ \frac{r_c}{a[\eta_{r,\text{ac}}]r}, \, \frac{\tau_c}{a[\eta_{r,\text{ac}}]r} \right] \bigg) \notag\\
& +  \mathcal S_{ij}^{(\Psi|\text{tail})}[\eta,\vec x]  \bigg( 1 + \mathcal O \left[ \frac{r_c}{a[\eta_{r,\text{ac}}]r}, \,  \frac{r_{c*}}{\Delta t} \right] \bigg) \Bigg\} ,
\label{Weyl_ScalarContribution_4DRadiation_FarZone}
\end{align}
with the leading acoustic direct $\mathcal S_{ij}^{(\Psi|\text{direct})}$ and tail $\mathcal S_{ij}^{(\Psi|\text{tail})}$ portions defined by
\begin{align}
\mathcal S_{ij}^{(\Psi|\text{direct})}[\eta,\vec x]  \equiv \,&  \big(\delta_{ij}-3\widehat r_i\widehat r_j\big) \frac{4 G_\text N}{r \eta^2}  \int_{\mathbb{R}^{3}} \dd^{3} \vec x' \, \left( {}^{(\text{a})}  T_{00}\big[\eta - \sqrt 3 \big( r - \vec x' \cdot \widehat r \big) ,\vec x'\big] \right. \notag\\
& \hspace{6cm} + \left. {}^{(\text{a})}  T_{ll} \big[\eta - \sqrt 3 \big( r - \vec x' \cdot \widehat r \big) ,\vec x'\big] \right)  ,
\label{AcousticConeWeyl_4DRadiation_FarZone}
\end{align}
\begin{align}
\mathcal S_{ij}^{(\Psi|\text{tail})}[\eta,\vec x]  \equiv \,& \big(\delta_{ij}-3\widehat r_i\widehat r_j\big) \frac{4 G_\text N}{3 r^3}   \int_{\mathbb{R}^{3}} \dd^{3} \vec x'  \int_0^{\eta - \sqrt 3 r - 0^+}  \dd\eta' \,\frac{(\eta^3 - \eta'^3)}{  \eta^3 \eta'} \notag\\
& \hspace{6cm} \times \left({}^{(\text{a})}  T_{00}[\eta',\vec x'] + {}^{(\text{a})}  T_{ll}[\eta',\vec x'] \right) ,
\label{AcousticTailWeyl_4DRadiation_FarZone}
\end{align}
respectively, where the (acoustic) retarded time $\eta_{r,\text{ac}} \equiv \eta - \sqrt 3 r$ at which $r_c$ and $\tau_c$ are both evaluated, while the $r_{c*}$ is the source's size at the peak time $\eta_*$, and the suppression factor $r_{c*}/\Delta t$ corresponds to the ratio of the additional acoustic-cone piece coming from Taylor expanding the exact tail portion of eq.~\eqref{Weyl_ScalarContribution_4DRadiation} to the leading tail effect of the latter itself, namely eq.~\eqref{AcousticTailWeyl_4DRadiation_FarZone}.

Similarly, in the non-relativistic limit as $r_c/\tau_c \to 0$, we may re-write these far-zone expressions in terms of the mass monopole and quadrupole moments defined in eqs.~\eqref{TotalMass_Cosmology} and \eqref{QuadrupoleMoment_Cosmology}, respectively,
{\allowdisplaybreaks
\begin{align}
\hspace{-1em}\mathcal S_{ij}^{(\Psi|\text{direct})}[\eta,\vec x] & \approx  \big(\delta_{ij}-3\widehat r_i\widehat r_j\big)  \frac{4 G_\text N}{r \eta^2 }  \frac{\eta_0}{\big(\eta - \sqrt 3 r \big)}  \Bigg\{ {}^{(\text{a})} M \big[\eta - \sqrt 3 r\big] + \frac12 \left( \frac{\eta_0}{\eta - \sqrt 3 r} \right)^2 {}^{(\text{a})} \ddot I_{ll}\big[\eta - \sqrt 3 r\big] \Bigg\} ,
\label{AcousticConeWeyl_4DRadiation_FarZone_Multipole}\\
\hspace{-1em}\mathcal S_{ij}^{(\Psi|\text{tail})}[\eta,\vec x] & \approx  \big(\delta_{ij}-3\widehat r_i\widehat r_j\big)  \frac{4G_\text N}{3r^3} \int_0^{\eta - \sqrt 3 r - 0^+}  \dd\eta' \,\frac{\eta_0(\eta^3 - \eta'^3)}{  \eta^3 \eta'^2} \Bigg\{  {}^{(\text{a})}M[\eta']  +  \frac12 \left( \frac{\eta_0}{\eta'} \right)^2 {}^{(\text{a})} \ddot I_{ll}[\eta'] \Bigg\},
\label{AcousticTailWeyl_4DRadiation_FarZone_Multipole}
\end{align}}%
where ${}^{(\text{a})}I_{ll} \equiv \delta^{ij} {}^{(\text{a})} I_{ij}$ and the first-order conservation laws \eqref{Coserve1_Leading} and \eqref{Coserve2_Leading} have been employed.

As already alluded to in the previous section, the acoustic tail effect, in the far-field regime, is not always suppressed in comparison with its acoustic-cone counterpart, as can be seen in the following acoustic tail-to-cone ratio,
\begin{align}
\left|\frac{\delta_1 C^{(\Psi|\text{tail})}[\eta,\vec x]}{\delta_1 C^{(\Psi|\text{direct})}[\eta,\vec x] } \right| \sim \frac{\Delta t}{a[\eta_*]r} \cdot \frac1{H[\eta]\big( a[\eta]r\big) } \cdot \frac{a[\eta]}{a[\eta_*]}  \left( 1 - \left( \frac{a[\eta_*]}{a[\eta]}\right)^3 \right) ,
\label{AcousticTailToCone_4DRadiation}
\end{align}
whose amplitude, in fact, depends on the hierarchy of the scales involved. The minimum tail-to-direct ratio, to be of the order $\Delta t/(a[\eta_*]r)$, is reached when the peak production of GWs happens near the retarded time $\eta_{r,\text{ac}}$, i.e., $\eta_*\approx \eta_{r,\text{ac}}$. In this scenario, if the physical observer-source distance at peak time, $a[\eta_*]r$, is sufficiently greater than the duration of the GW source, $\Delta t $, then the acoustic direct signals could still dominate over their tail counterpart. On the other hand, if the active period of GW production occurs fairly early in the past, as long as the factor $\Delta t/(a[\eta_*]r)$ is not as small as the order of $H[\eta](a[\eta]r)$, the acoustic tail signals could actually comprise a large proportion of the scalar tidal forces.

\subsection{Far-Zone JWKB Limit: Space Distortions and Polarization Patterns}
\label{JWKB_Polarization}

Having studied the high frequency limit of the traceless geometric tidal forces in a radiation dominated universe, let us now turn to a closely related issue. What is the corresponding `JWKB limit' of the distortion of space driven by gravitational radiation, as experienced by free-falling co-moving test masses sprinkled within such a cosmology; and quantified via eq.~\eqref{FractionalDistortion}? Of course, the spin-2 polarization patterns are well known in a Minkowski background. We shall not only extend these tensor results to the $w=1/3$ cosmological case -- but also uncover the acoustic-gravitational ones, which have no counterpart in flat spacetime, de Sitter, nor in matter dominated $w=0$ universes.

{\bf Synchronous-Weyl Relation} \qquad The key object within the fractional distortion formula of eq.~\eqref{FractionalDistortion} is the synchronous gauge metric perturbation $\chi^{(s)}_{ij}$. Let us first explain why, the dominant term of $\delta_1 C^i_{\phantom{i}0j0}$ is in fact its acceleration; namely
\begin{align}
\delta_1 C^i{}_{0j0}& \approx   -\frac12 \ddot\chi^{(s)}_{ij} .
\label{WeylSynchMetricPerturbation_LeadingOrder}
\end{align}
Since the linearized Weyl tensor is gauge invariant, in the high frequency $\omega$ limit, we may then proceed to use the JWKB results for the Weyl tensor components obtained in \S \eqref{JWKB_Tidal} above, to solve for $\chi^{(s)}_{ij}$ -- i.e., by equating the dominant contributions to Weyl from equations \eqref{WeylTensor_4DRadiationDomination_FarZone} and \eqref{Weyl_ScalarContribution_4DRadiation_FarZone} to the right hand side of eq.~\eqref{WeylSynchMetricPerturbation_LeadingOrder}.

We begin with the ``on-shell'' relationship between the Riemann and Weyl tensors,
\begin{align}
C^{\rho}{}_{\sigma\mu\nu}=\,& R^{\rho}{}_{\sigma\mu\nu} - 8\pi G_\text N \left( \delta^\rho_{[\mu}T_{\nu]\sigma}-g_{\sigma[\mu}T_{\nu]}{}^\rho  - \frac{2}{3} \delta^\rho_{[\mu}g_{\nu]\sigma} g^{\alpha\beta}T_{\alpha\beta} \right) ;
\label{RiemannWeyl}
\end{align}
where, on the right hand side, the Einstein's equation $G_{\mu\nu} = 8\pi \GN T_{\mu\nu}$ has been imposed on the trace parts of the Riemann tensor, with $T_{\mu\nu}$ referring to the energy-momentum tensor of the total matter content -- both the perfect fluid and the isolated astrophysical system. Exploiting the $a^2 (\eta_{\mu\nu} + \chi_{\mu\nu})$ form of our cosmological geometry, we perform a conformal transformation of the Riemann tensor to reveal, in the synchronous gauge,
\begin{align}
R^i{}_{0j0} = - \delta_{ij}  \dot{\mathcal H} -\frac12 \ddot\chi^{(s)}_{ij}  +  \frac12 \mathcal H  \dot \chi^{(s)}_{ij} + \mathcal{O}\left[ \left(\chi_{mn}^{(s)}\right)^2 \right] .
\label{RiemannToSynchronousMetric}
\end{align}
At the background level, $\mathcal H$ is governed by the Friedmann equations,
{\allowdisplaybreaks
	\begin{align}
	\mathcal H^2     &=    \frac{8 \pi G_\text N}{3}  \overline T_{00},
	\label{FriedmannEq_1}   \\
	\dot{\mathcal H} &=  -\frac{4\pi G_\text N}{3}  \left(\overline T_{00} + 3 a^2 \overline p \vphantom{\dot A}\right),
	\label{FriedmannEq_2}
	\end{align}}%
where $\overline T_{\mu\nu}$, being the zeroth-order total stress tensor, involving only the background perfect fluid, so it takes a diagonal form with isotropic pressure $\overline T_{ij} \equiv \delta_{ij} a^2 \overline{p}$. Plugging eqs.~\eqref{RiemannToSynchronousMetric}, \eqref{FriedmannEq_1}, and \eqref{FriedmannEq_2} into eq.~\eqref{RiemannWeyl}, one would now obtain, up to first order in ${\chi}_{ij}^{(s)}$ and $\delta_1 T_{\mu\nu}$:
{\allowdisplaybreaks
	\begin{align}
	\delta_1 C^i{}_{0j0}=\,&   -\frac12 \ddot\chi^{(s)}_{ij}  +  \frac12 \mathcal H \dot \chi^{(s)}_{ij} \nonumber\\
	&+ 4\pi G_\text N \bigg\{  \left(  \delta_1T_{ij}  -  a^2 \overline p \chi_{ij}^{(s)}  \right)
	-  \frac{\delta_{ij}}3  \bigg( \delta_1T_{00}   +  2 \left(         \delta_1T_{ll}  -  a^2 \overline{p}   \chi_{ll}^{(s)}  \right)  \bigg)  \bigg\}      .
	\label{WeylSynchMetricPerturbation_OnShell}
	\end{align}}%
Notice from eq.~\eqref{WeylSynchMetricPerturbation_OnShell} that the linear-order piece of $T_{\mu\nu}$, denoted by $\delta_1 T_{\mu\nu}$, consists not just of the compact astrophysical sources, i.e., ${}^{(\text{a})}T_{\mu\nu}$, but also of the first-order perturbations of the fluid that drives the cosmic expansion. In other words, even at linear order in perturbations, the precise connection between the synchronous gauge gravitational perturbation and that of the Weyl tensor components requires not only understanding gravitational dynamics; but those of the first order perturbed fluid as well.

Nonetheless, let us argue that the first term on the right hand side of eq.~\eqref{WeylSynchMetricPerturbation_OnShell} is the dominant one -- i.e., eq.~\eqref{WeylSynchMetricPerturbation_LeadingOrder} is justified -- as long as the characteristic timescale of $\chi^{(s)}_{ij}$, which in turn is associated with that of the GW source(s), is much smaller than the age of the universe. For, we may estimate $\ddot\chi^{(s)}_{ij}$, $\mathcal H  \dot \chi^{(s)}_{ij}$, and $ G_\text N \delta_1T_{\mu\nu}$ in eq.~\eqref{WeylSynchMetricPerturbation_OnShell}, respectively, to scale as
{\allowdisplaybreaks
	\begin{align}
	\ddot \chi^{(s)}_{ij} &\sim  \tau_{\chi}^{-2} a^2 \chi^{(s)}_{ij} \left( 1 + \mathcal O[H\tau_\chi] \vphantom{\dot A}\right) ,
	\label{DoubleDotSynchronousMetric_WeylSynchronous} \\
	\mathcal H \dot \chi^{(s)}_{ij} &\sim  H  \tau_\chi^{-1} a^2 \chi^{(s)}_{ij} ,
	\label{SingleDotSynchronousMetric_WeylSynchronous}\\
	G_\text N \delta_1 T_{\mu\nu}   &\sim  G_\text N \left( \overline T \chi^{(s)} \right)_{\mu\nu} \sim H^2 a^2  \left( \chi^{(s)} \right)_{\mu\nu} ;
	\label{LinearizedStressTensor_WeylSynchronous}
	\end{align}}%
where $\tau_\chi$ denotes the timescale of the synchronous-gauge perturbations in terms of the cosmic time $t = \int \dd \eta \, a[\eta]$, and $\delta_1T_{\mu\nu} \sim \left( \overline T \chi^{(s)} \right)_{\mu\nu}$. As the Hubble parameter is also inversely related to the age of the universe, the factor $H\tau_\chi$ appears to be a small ratio of the two scales, i.e., $H\tau_\chi \ll 1$, indicating eqs.~\eqref{SingleDotSynchronousMetric_WeylSynchronous} and \eqref{LinearizedStressTensor_WeylSynchronous} are both $H\tau_\chi$ suppressed relative to eq.~\eqref{DoubleDotSynchronousMetric_WeylSynchronous}. To sum, we have arrived at the estimate:
\begin{align}
\delta_1 C^i{}_{0j0}=\,&   -\frac12 \ddot\chi^{(s)}_{ij} \left( 1 + \mathcal O[H\tau_\chi] \vphantom{\dot A}\right) .
\end{align}
From eq.~\eqref{RiemannToSynchronousMetric}, we may in turn infer that
\begin{align}
\delta_1 C^i{}_{0j0} \approx \delta_1R^i{}_{0j0} .
\end{align}
Physically speaking, the Weyl components $\delta_1 C^i{}_{0j0}$ provide the dominant contributions to the first-order tidal forces.

{\bf Monochromatic waves} \qquad To identify the radiative behaviors of eqs.~\eqref{WeylTensor_4DRadiationDomination_FarZone} and \eqref{Weyl_ScalarContribution_4DRadiation_FarZone}, and characterize their oscillatory polarization patterns, we shall now focus on the monochromatic component waves associated with these far-zone solutions. This also serves as a practical approximation to the observed signals, for realistic GW detectors are only sensitive to a limited range of frequencies. However, unlike the Minkowski spin-2 waveforms, the frequency-Fourier transform cannot be directly exploited to decompose eqs.~\eqref{WeylTensor_4DRadiationDomination_FarZone} and \eqref{Weyl_ScalarContribution_4DRadiation_FarZone} into their individual frequency modes, due to the overall time-dependent amplitudes, as well as the constraint for conformal time to be strictly positive $\eta>0$\footnote{Here, we are primarily interested in the propagating monochromatic waves in ``frequency space''. This is not be to confused with the spatial Fourier transform commonly exploited in the cosmology literature, in which some cosmological observables are expressible in terms of the Fourier modes.
}. Nevertheless, at high frequencies, we may instead Fourier decompose the $\delta$-functions encapsulated within eqs.~\eqref{ttGraviton_4DRadiation} and \eqref{AcousticConeWeyl_4DRadiation_FarZone}, the direct parts of the signals, to have them re-expressed in terms of the superpositions of the outgoing JWKB spherical waves propagating on the null and acoustic cones, respectively,
{\allowdisplaybreaks
\begin{align}
\chi^{(\text{tt}|\text{4D rad})}_{ij}[\eta,\vec x]  \approx \,&  P^{(\text{4D})}_{ijmn} \frac{8 G_\text N}{r\eta} \text{Re} \bigg[ \int_0^\infty \frac{\dd \omega}{2\pi} \, {}^{(\text{a})}\widetilde{\overline{\mathcal T}}_{mn}[\omega,\omega \widehat r]  e^{-i\omega(\eta - r)} \bigg] ,
\label{ttGraviton_4DRadiation_FrequencyDecomposition}  \\
\mathcal S_{ij}^{(\Psi|\text{direct})}[\eta,\vec x]  = \,& \big(\delta_{ij}-3\widehat r_i\widehat r_j\big) \frac{8 G_\text N}{r \eta^2} \text{Re} \bigg[  \int_0^\infty \frac{\dd\omega}{2\pi} \,
{}^{(\text{a})} \widetilde{\mathcal T}\big[\omega , \sqrt 3 \omega \widehat r\big]  e^{-i\omega \left(\eta  - \sqrt 3 r \right) } \bigg] ,
\label{AcousticConeWeyl_4DRadiation_FrequencyDecomposition}
\end{align}}%
where ${}^{(\text{a})}\widetilde{\overline{\mathcal T}}_{mn}$ and ${}^{(\text{a})} \widetilde{\mathcal T}$ are defined respectively as
{\allowdisplaybreaks
\begin{align}
{}^{(\text{a})}\widetilde{\overline{\mathcal T}}_{mn}[\omega,\omega \widehat r] & \equiv \int_0^\infty \dd \eta' \,\eta' \int_{\mathbb{R}^{3}} \dd^{3}\vec x'  \,  e^{i\omega\left( \eta'  -  \vec x' \cdot \widehat r \right)} {}^{(\text{a})}T_{mn}[\eta',\vec x'] ,
\label{FrequencyTransformOfSourse_Spin2_4DRadiation} \\
{}^{(\text{a})}\widetilde{\mathcal T}\big[\omega , \sqrt 3\omega \widehat r\big] & \equiv \int_0^\infty \dd \eta' \int_{\mathbb{R}^{3}} \dd^{3}\vec x'  \, e^{i\omega\left( \eta'  - \sqrt 3 \vec x' \cdot \widehat r \right)} \left( {}^{(\text{a})}T_{00}[\eta',\vec x'] + {}^{(\text{a})} T_{ll}[\eta',\vec x'] \right) .
\label{FrequencyTransformOfSourse_AcousticCone_4DRadiation}
\end{align}}%
For each $\omega$-mode in eqs.~\eqref{ttGraviton_4DRadiation_FrequencyDecomposition} and \eqref{AcousticConeWeyl_4DRadiation_FrequencyDecomposition}, denoted by
{\allowdisplaybreaks
\begin{align}
\widetilde \chi^{(\text{tt}|\text{4D rad})}_{ij}[\eta,\vec x]  \equiv \,&  P^{(\text{4D})}_{ijmn} \frac{8 G_\text N}{r\eta} \text{Re} \left[ {}^{(\text{a})}\widetilde{\overline{\mathcal T}}_{mn}[\omega,\omega \widehat r]  e^{-i\omega(\eta - r)} \right] ,
\label{ttGraviton_4DRadiation_SingleFrequency}  \\
\widetilde{\mathcal S}_{ij}^{(\Psi|\text{direct})}[\eta,\vec x]  \equiv \,& \big(\delta_{ij}-3\widehat r_i\widehat r_j\big) \frac{8 G_\text N}{r \eta^2} \text{Re} \left[
{}^{(\text{a})} \widetilde{\mathcal T}\big[\omega , \sqrt 3\omega \widehat r\big] e^{-i\omega \left(\eta  - \sqrt 3 r \right) } \right] ,
\label{AcousticConeWeyl_4DRadiation_SingleFrequency}
\end{align}}%
$\omega$ is physically related to the GW frequency $\omega_{\text{gw}}$, as measured by a co-moving observer, via the redshift relationship $\omega_{\text{gw}}[\eta] = \omega/a[\eta]$ in the high-frequency JWKB limit. Furthermore, with $\omega_{\text{gw}}[\eta_r] \sim \tau_c^{-1}$ identified for the null propagation, the far-zone condition $\tau_c/(a[\eta_r]r) \ll 1$, together with the suppression factor $H[\eta_r]\tau_c \ll 1$, will translate into the limits $\omega r \sim a[\eta_r]r/\tau_c\gg 1$ and $\omega \eta \sim 1/(H[\eta_r]\tau_c)\gg 1$ in ``frequency space'', which applies to its acoustic counterpart as well.

In terms of these monochromatic JWKB waves, i.e., eqs.~\eqref{ttGraviton_4DRadiation_SingleFrequency} and \eqref{AcousticConeWeyl_4DRadiation_SingleFrequency}, the dominant (spin-2) GW tidal forces now read (cf.~eq.~\eqref{WeylTensor_4DRadiationDomination_FarZone})
\begin{align}
\delta_1  C^{(\text{4D rad}|\omega)i}{}_{0j0} [\eta,\vec x] & \equiv  -\frac12 \ddot{\widetilde \chi}^{(\text{tt}|\text{4D rad})}_{ij}[\eta,\vec x]  , \notag\\
& \approx P^{(\text{4D})}_{ijmn} \frac{4 G_\text N}{r\eta} \omega^2 \, \text{Re}\left[ {}^{(\text{a})}\widetilde{\overline{\mathcal T}}_{mn}[\omega,\omega \widehat r]  e^{-i\omega(\eta - r)} \right] ,
\label{Spin2WeylTensor_SingleFrequencyMode}
\end{align}
whereas the direct portion of the leading scalar ones \eqref{Weyl_ScalarContribution_4DRadiation_FarZone} gives
\begin{align}
\delta_1 C^{(\Psi,\text{direct}|\text{4D rad}|\omega)i}{}_{0j0}[\eta,\vec x] & \equiv - \frac12 \widetilde{\mathcal S}_{ij}^{(\Psi|\text{direct})}[\eta,\vec x] , \notag\\
& = \big(3\widehat r_i\widehat r_j - \delta_{ij}\big) \frac{4 G_\text N}{r \eta^2} \text{Re}\left[ {}^{(\text{a})} \widetilde{\mathcal T}\big[\omega , \sqrt 3\omega \widehat r\big]   e^{-i\omega \left(\eta  - \sqrt 3 r \right) } \right].
\label{ScalarWeylTensor_SingleFrequencyMode}
\end{align}
In these expressions, the $(H\tau_c)^2$ suppression in amplitude found in eq.~\eqref{AcousticToNull_Direct_4DRadiation}, can roughly be accounted for by a factor of $1/(\omega \eta)$ in their prefactors, with another $1/(\omega \eta')$ within the integrands of ${}^{(\text{a})} \widetilde{\mathcal T}$ and $\omega {}^{(\text{a})}\widetilde{\overline{\mathcal T}}_{mn}$ in terms of eqs.~\eqref{FrequencyTransformOfSourse_Spin2_4DRadiation} and \eqref{FrequencyTransformOfSourse_AcousticCone_4DRadiation}.

{\bf Matching of Inhomogeneous/Homogeneous Solutions} \qquad  If eqs.~\eqref{Spin2WeylTensor_SingleFrequencyMode} and \eqref{ScalarWeylTensor_SingleFrequencyMode} were both considered to be dynamical propagating waves, one would expect them to coincide with their homogeneous counterparts in the far-field regime. The exact plane-wave-like solution of the linearized Weyl tensor, denoted by $\delta_1 C^{(\text{PW})i}{}_{0j0}$, can be readily obtained for all $d\geq4$ by solving the source-less counterparts of eqs.~\eqref{Spin-2_WaveEquation_Cosmology} and \eqref{Psi_WaveEq_Cosmology} for $D_{ij}$ and $\Psi$, respectively, followed by inserting them and $\Phi = (d-3) \Psi$ into eq.~\eqref{LinearizedWeyl_GaugeInvariant},
\begin{align}
\delta_1 C^{(\text{PW})i}{}_{0j0}[\eta,\vec x] & =  k^2 (k\eta)^{\frac12 -\frac{d-2}{q_w}} \text{Re} \left[\epsilon_{ij}[\vec k]  \bigg(  H^{(2)}_{\frac32 + \frac{d - 2}{ q_w}}[k\eta] - \frac1{k\eta} H^{(2)}_{\frac12 + \frac{d - 2}{ q_w}}[k\eta]  - \frac1{d-3} H^{(2)}_{-\frac12 + \frac{d - 2}{ q_w}}[k\eta]  \bigg) e^{i \vec k  \cdot \vec x } \right]  \notag\\
& +  \left(  (d-1)\widehat k_i \widehat k_j  -  \delta_{ij} \right) k^2  (k\eta)^{-\frac12 -\frac{d - 2}{ q_w}}  \text{Re} \left[ b[\vec k]  H^{(2)}_{\frac12 + \frac{d - 2}{ q_w}}[k\eta]e^{ i \vec k \cdot \frac{\vec x}{\sqrt w}} \right]   , \qquad k \equiv |\vec k| .
\label{Weyl_HomogeneousPW_AllDimensions}
\end{align}
The unit vector $\widehat k \equiv \vec k/|\vec k|$ points towards the direction of the wave propagation; $H^{(2)}_\nu$ is the Hankel function of the second kind; the spin-2 polarization tensor $\epsilon_{ij}[\vec k]$ obeys the traceless-transverse constraints $\delta^{ij}\epsilon_{ij} = 0 = k^i \epsilon_{ij}$; whereas the scalar amplitude $b[\vec k]$ remains arbitrary. Note that the first line of eq.~\eqref{Weyl_HomogeneousPW_AllDimensions} comes from the spin-2 gravitons $D_{ij}$, while its second line is due to the Bardeen scalar potentials $\Psi$ and $\Phi$; the vector mode $V_i$ does not contribute at all.

Within the JWKB approximation, where the GW wavelength $\lambda_{\text{gw}} = 2\pi a/k \ll H^{-1}$, or equivalently $k \eta \gg 1$, we may employ the asymptotic expansion of $H_\nu^{(2)}[k\eta]$ for large arguments,
\begin{align}
H_\nu^{(2)}[k\eta] = \sqrt{\frac2{\pi k\eta}} \, e^{- i\left( k\eta - \frac{\nu \pi}2 - \frac{\pi}4 \right)} \left( 1 + \mathcal O \left[\frac1{k\eta}\right]\right) ,
\end{align}
to extract the asymptotic behavior of eq.~\eqref{Weyl_HomogeneousPW_AllDimensions}:
\begin{align}
\delta_1 C^{(\text{PW})i}{}_{0j0}[\eta,\vec x] =\,& \left( \delta_1 C^{(\text{PW}|g)i}{}_{0j0}[\eta,\vec x]  +  \delta_1 C^{(\text{PW}|\Psi)i}{}_{0j0}[\eta,\vec x] \right) \left( 1 + \mathcal O\left[\frac1{k\eta}\right] \right) .
\label{Weyl_HomogeneousPW_AllDimensions_WKB}
\end{align}
The spin-2-only $\delta_1 C^{(\text{PW}|g)i}{}_{0j0}$ and spin-0-only $\delta_1 C^{(\text{PW}|\Psi)i}{}_{0j0}$ tidal forces now denote their leading order expressions,
{\allowdisplaybreaks
\begin{align}
\delta_1 C^{(\text{PW}|g)i}{}_{0j0}[\eta,\vec x] & \equiv  k^2 (k\eta)^{-\frac{d-2}{q_w}} \text{Re} \left[\epsilon_{ij}[\vec k] e^{-i k\left( \eta - \widehat k \cdot \vec x \right)} \right]  ,
\label{WeylHomogeneousPW_AllDimensions_WKB_Spin2}  \\
\delta_1 C^{(\text{PW}|\Psi)i}{}_{0j0}[\eta,\vec x] & \equiv   \left( (d-1)\widehat k_i \widehat k_j - \delta_{ij} \right) k^2 (k\eta)^{-1 - \frac{ d - 2}{ q_w}} \text{Re} \left[ b[\vec k] e^{-i k\left( \eta - \widehat k \cdot \frac{\vec x}{\sqrt w} \right)} \right] ;
\label{WeylHomogeneousPW_AllDimensions_WKB_BardeenScalars}
\end{align}}%
with all the numerical coefficients absorbed into re-definitions of $\epsilon_{ij}[\vec k]$ and $b[\vec k]$. We observe from eqs.~\eqref{WeylHomogeneousPW_AllDimensions_WKB_Spin2} and \eqref{WeylHomogeneousPW_AllDimensions_WKB_BardeenScalars} that, apart from the undetermined amplitudes $\epsilon_{ij}[\vec k]$ and $b[\vec k]$, the prefactor of the latter is already $1/(k\eta)$ suppressed relative to that of the former for all relevant spacetime dimensions.

When we specialize to 4D radiation domination, the leading spin-2 null waves \eqref{WeylHomogeneousPW_AllDimensions_WKB_Spin2} and the scalar acoustic ones \eqref{WeylHomogeneousPW_AllDimensions_WKB_BardeenScalars} are, respectively,
{\allowdisplaybreaks
\begin{align}
\delta_1 C^{(\text{PW}|g|\text{4D rad})i}{}_{0j0}[\eta,\vec x] & =  \frac k\eta \, \text{Re} \left[\epsilon_{ij}[\vec k] e^{-i k\left( \eta - \widehat k \cdot \vec x \right)} \right]  ,
\label{WeylHomogeneousPW_4DRadiation_WKB_Spin2}  \\
\delta_1 C^{(\text{PW}|\Psi|\text{4D rad})i}{}_{0j0}[\eta,\vec x] & =  \left(  3\widehat k_i \widehat k_j - \delta_{ij} \right) \frac1{\eta^2} \, \text{Re} \left[ b[\vec k] e^{-i k\left( \eta - \sqrt 3 \, \widehat k \cdot \vec x \right)} \right] .
\label{WeylHomogeneousPW_4DRadiation_WKB_BardeenScalars}
\end{align}}%
Comparing these results with their inhomogeneous counterparts \eqref{Spin2WeylTensor_SingleFrequencyMode} and \eqref{ScalarWeylTensor_SingleFrequencyMode}, we find that the matching of both the null/acoustic far-zone JWKB waveforms can indeed be established by the following correspondence between the two sides:
{\allowdisplaybreaks
\begin{align}
\vec k & \leftrightarrow \omega \widehat r , \\
\epsilon_{ij}[\vec k] &  \leftrightarrow  \frac{4 G_\text N }r P^{(\text{4D})}_{ijmn} \, \omega {}^{(\text{a})}\widetilde{\overline{\mathcal T}}_{mn}[\omega,\omega \widehat r]  ,
\label{Spin2PolarizationTensor_Matching}\\
b[\vec k]  &   \leftrightarrow \frac{4 G_\text N}{r }   {}^{(\text{a})} \widetilde{\mathcal T}\big[\omega , \sqrt 3\omega \widehat r\big]  .
\label{BardeenScalarAmplitude_Matching}
\end{align}}%
In other words, the remaining free parameters in the homogeneous solutions \eqref{WeylHomogeneousPW_4DRadiation_WKB_Spin2} and \eqref{WeylHomogeneousPW_4DRadiation_WKB_BardeenScalars} can, in fact, be fixed through this matching procedure in eqs.~\eqref{Spin2PolarizationTensor_Matching} and \eqref{BardeenScalarAmplitude_Matching}, and furthermore, it is the acoustic-cone part of the scalar tidal forces, instead of the acoustic tails \eqref{AcousticTailWeyl_4DRadiation_FarZone}, that agrees with the JWKB plane waves in the far zone, despite the potentially larger magnitude of the latter (recall eq.~\eqref{AcousticTailToCone_4DRadiation}). On physical grounds, this identification does further support the assertion that the Bradeen scalars, like the spin-2 TT gravitons, should be regarded as part of gravitational radiation in cosmology.

{\bf Gravitational Polarization Patterns} \qquad  Finally, let us now study the polarization patterns tidally-induced by the gravitational tensor and scalar radiation. From eqs.~\eqref{WeylSynchMetricPerturbation_LeadingOrder}, \eqref{Spin2WeylTensor_SingleFrequencyMode}, and \eqref{ScalarWeylTensor_SingleFrequencyMode}, we may solve for $\chi^{(s)}_{ij}$ required in fractional distortion formula of eq.~\eqref{FractionalDistortion} within the high-frequency limit, by first splitting it into the spin-2 and scalar sectors,
\begin{align}
\chi^{(s)}_{ij} = \chi^{(s|g)}_{ij} + \chi^{(s|\Psi)}_{ij} ,
\end{align}
with the ansatz that the former takes the same JWKB form as $\widetilde \chi^{(\text{tt}|\text{4D rad})}_{ij}$ in eq.~\eqref{ttGraviton_4DRadiation_SingleFrequency} and the latter as $\widetilde{\mathcal S}_{ij}^{(\Psi|\text{direct})}$ in eq.~\eqref{AcousticConeWeyl_4DRadiation_SingleFrequency}, followed by equating their accelerations to eqs.~\eqref{Spin2WeylTensor_SingleFrequencyMode} and \eqref{ScalarWeylTensor_SingleFrequencyMode}, which then reveals that, at high frequencies,
\begin{align}
\chi^{(s|g)}_{ij}[\eta,\vec x] & \approx \widetilde \chi^{(\text{tt}|\text{4D rad})}_{ij}[\eta,\vec x] \notag\\
&  = P^{(\text{4D})}_{ijmn} \frac{8 G_\text N}{r\eta} \text{Re} \left[ {}^{(\text{a})}\widetilde{\overline{\mathcal T}}_{mn}[\omega,\omega \widehat r]  e^{-i\omega(\eta - r)} \right] ,
\label{SynchMetricPerturbations_SingleFrequency_Spin2}
\end{align}
\begin{align}
\chi^{(s|\Psi)}_{ij}[\eta, \vec x]  & \approx  - \frac1{\omega^2} \widetilde{\mathcal S}_{ij}^{(\Psi|\text{direct})}[\eta,\vec x] \notag\\
& = \big( 3\widehat r_i\widehat r_j - \delta_{ij} \big) \frac{8 G_\text N}{r (\omega\eta)^2} \text{Re} \left[
{}^{(\text{a})} \widetilde{\mathcal T}\big[\omega , \sqrt 3\omega \widehat r\big] e^{-i\omega \left(\eta  - \sqrt 3 r \right) } \right] .
\label{SynchMetricPerturbations_SingleFrequency_Scalar}
\end{align}
The amplitude of the scalar portion in eq.~\eqref{SynchMetricPerturbations_SingleFrequency_Scalar} is likewise suppressed compared to that of its spin-2 counterpart \eqref{SynchMetricPerturbations_SingleFrequency_Spin2}.

If the GW wavelength is sufficiently larger than the proper size of the hypothetical GW detector employed to probe the cosmological tidal distortion, we may drop the integral over $\lambda$ in eq.~\eqref{FractionalDistortion} as the perturbations \eqref{SynchMetricPerturbations_SingleFrequency_Spin2} and \eqref{SynchMetricPerturbations_SingleFrequency_Scalar} are approximately constant from one end to the other. Then, with the unit vector $\widehat n$ in eq.~\eqref{FractionalDistortion} parameterized in terms of the spherical coordinates as follows,
\begin{align}
\widehat n^i = \sin[\theta] \cos[\phi] \widehat e_x^i  +  \sin[\theta] \sin[\phi] \widehat e_y^i  + \cos[\theta] \widehat r^i,
\label{UnitVector_Parameterization}
\end{align}
where $\widehat e_x$ and $\widehat e_y$ are the mutually orthogonal unit vectors lying on the two-dimensional spatial plane perpendicular to the radial direction $\widehat r$, along which the wave is propagating, the strain \eqref{FractionalDistortion} generated by the spin-2 waves of eq.~\eqref{SynchMetricPerturbations_SingleFrequency_Spin2} is given by
\begin{align}
\left(\frac{\delta L}{L_0}\right)_{\text{spin-2} } & \approx \frac12 \widehat n^i\widehat n^j \chi^{(s|g)}_{ij} \notag\\
& =   \frac{\sin^2\theta}2 \, \text{Re} \left[ \left( \vphantom{\dot A} h_+[\omega,\omega \widehat r] \cos[2\phi]  +   h_\times[\omega,\omega \widehat r] \sin[2\phi]   \right) e^{-i\omega \left(\eta - r \right)} \right] ,
\label{Strain_Spin2GWs}
\end{align}
where $L_0$ denotes the original proper distance between two test masses before a GW impinges on the detector, and the two independent polarizations $h_+$ and $h_\times$ are respectively defined as
{\allowdisplaybreaks
\begin{align}
h_+ [\omega,\omega \widehat r]& \equiv \frac{4 G_\text N}{r\eta} \left( {}^{(\text{a})}\widetilde{\overline{\mathcal T}}_{xx}[\omega,\omega \widehat r] - {}^{(\text{a})}\widetilde{\overline{\mathcal T}}_{yy}[\omega,\omega \widehat r] \right) , \\
h_\times[\omega,\omega \widehat r] & \equiv \frac{8 G_\text N}{r\eta} {}^{(\text{a})}\widetilde{\overline{\mathcal T}}_{xy}[\omega,\omega \widehat r]  ,
\end{align}}%
with $ {}^{(\text{a})}\widetilde{\overline{\mathcal T}}_{AB} \equiv \widehat e_{A}^i \widehat e_{B}^j {}^{(\text{a})}\widetilde{\overline{\mathcal T}}_{ij} $, $A,B \in \{x,y\}$. The result \eqref{Strain_Spin2GWs} clearly demonstrates the familiar spin-2 polarization pattern, which apart from the redshift factor $1/a \propto 1/\eta$, is very similar to its Minkowski counterpart. Specifically, the proper spatial displacement between the test masses will not be affected when the pair are aligned with the wave propagation, i.e., $\theta = 0$, and the overall distortion amplitude reaches its maximum at $\theta = \pi/2$. Furthermore, for a fixed inclination angle $\theta \neq 0$, the maximum distortion of the $h_+$ polarization occurs at $\phi = 0$ and $\phi = \pi/2$, oscillating in a ``+'' shape, whereas the $h_\times$ type of polarization has the maximum distortion at $\phi = \pm \pi/4$ instead. One mode will coincide with the other under a rotation of $\pi/4$ about the radial direction.

The scalar-GW-induced strain is obtained by plugging eqs.~\eqref{SynchMetricPerturbations_SingleFrequency_Scalar} and \eqref{UnitVector_Parameterization} into eq.~\eqref{FractionalDistortion}, with the $\lambda$-integral discarded in the long-wavelength limit,
\begin{align}
\left(\frac{\delta L}{L_0}\right)_{\text{scalar}}  & \approx \frac12 \widehat n^i\widehat n^j \chi^{(s|\Psi)}_{ij} \notag\\
& = \frac{4 G_\text N}{r (\omega\eta)^2} \big( 3 \cos^2\theta  -  1 \big) \text{Re}\left[ {}^{(\text{a})} \widetilde{\mathcal T}\big[\omega , \sqrt 3\omega \widehat r\big]  e^{-i\omega \left(\eta  - \sqrt 3 r \right) } \right] .
\label{Strain_ScalarGWs}
\end{align}
This shows that the resulting polarization pattern, unlike the tensor one, is isotropic with respect to the azimuthal angle $\phi$; moreover, it reaches the maximum distortion when the pair of test masses are collinear with the wave propagation ($\theta = 0$) and remains undistorted at $\theta = \arccos[1/\sqrt 3\,]$. In other words, the scalar $\Psi$-waves would give rise to an extra but sub-leading ``longitudinal'' mode in the gravitational polarization patterns.

We also observe that, within the non-relativistic limit, the scalar polarization pattern \eqref{Strain_ScalarGWs} involves both the mass monopole and quadrupole moments of the GW source, revealed by eq.~\eqref{AcousticConeWeyl_4DRadiation_FarZone_Multipole} and the fact that $\widetilde{\mathcal S}_{ij}^{(\Psi|\text{direct})}$ in eq.~\eqref{SynchMetricPerturbations_SingleFrequency_Scalar} is the corresponding single ``frequency'' mode; whereas the spin-2 pattern \eqref{Strain_Spin2GWs} only involves the mass quadrupole moment, as can be seen in eqs.~\eqref{ttGraviton_4DRadiation_Quadrupole} and \eqref{SynchMetricPerturbations_SingleFrequency_Spin2}.

\section{Summary, Discussions, and Future Directions}
\label{Section_Summary}

The primary physical results of this paper are the far zone high frequency trace-free tidal forces induced by the acoustic-gravitational perturbations found in eq.~\eqref{Weyl_ScalarContribution_4DRadiation_FarZone}; as well as the corresponding fractional distortion of space in the freely-falling frame of co-moving observers, as encoded within the formula \eqref{Strain_ScalarGWs}. We highlight the scalar results over their tensor spin-2 cousins, because the latter are largely red-shifted Minkowski solutions whereas the former do not exist in the flat background, nor even in de Sitter $w=-1$ or matter dominated $w=0$ cosmologies.

Within the scalar-acoustic tidal forces, the astrophysical energy density ${}^{(\text{a})}T_{00}$ appears to source $\delta_1 C^{(\Psi)i}{}_{0j0}$ on an equal footing with the pressure term ${}^{(\text{a})} T_{ll}$. This suggests, isolated astrophysical systems may lose their mass through these acoustic-gravitational radiation -- a possibility already raised in \cite{Chu:2016ngc}. (Our results here are also consistent with the point made in \cite{Chu:2016ngc}; that mass loss will occur as long as the sum of the astrophysical source(s)' internal pressures is non-zero.) In the same vein, notice in the non-relativistic limit, scalar radiation involves both monopole and quadrupole moments (see eq.~\eqref{AcousticConeWeyl_4DRadiation_FarZone_Multipole}); whereas the tensor one only involves quadrupole one (see eq.~\eqref{ttGraviton_4DRadiation_Quadrupole}). Even though we are already calling these acoustic-gravitational perturbations `scalar radiation', to be certain they do indeed carry energy-momentum away from their emitter; we would have to embark on a nonlinear calculation of the quadratic piece of the Einstein-fluid equations, so as to extract the stress-energy (pseudo-)tensor of the gravitational perturbations. Perhaps this computation could shed light on the meaning of the acoustic tail versus cone terms; in particular, why the former appears to yield stronger signals than the latter in many circumstances.

Finally, the concrete results in this work focused exclusively on the radiation dominated $w = 1/3$ phase of our 4D universe. This is, of course, an important epoch; and we hope to further extend our analysis here by considering specific GW sources of potential physical relevance -- cosmic strings \cite{Damour:2000wa,Damour:2001bk} and binary primordial black holes, for instance. But we also wish to push our analytic understanding of cosmological gravitational waves to other relativistic equation-of-states $0 < w \leq 1$, and generalize the works of \cite{Ashtekar:2015lxa,Date:2016uzr,Bonga:2017dlx,Hoque:2017xop} in de Sitter to other constant-$w$ cosmologies.

\section{Acknowledgments}

YZC is supported by the Ministry of Science and Technology of the R.O.C. under the grant 106-2112-M-008-024-MY3. YZC thanks Beatrice Bonga, Wayne Hu, Glenn Starkman, and Richard Woodard for discussions. YWL is supported by the Ministry of Science and Technology of the R.O.C. under Project No.~MOST 108-2811-M-008-589. This work was also supported in part by the Ministry of Science and Technology of the R.O.C. under Project No.~MOST 108-2811-M-008-503.

\appendix

\section{Linearized Weyl Tensor in de Sitter and Matter-Dominated Universes} \label{LinearizedWeyl_deSitterAndMatterDomination}

Lying within the physical range of the constant equation-of-state $w$, $w=-1$ (de Sitter) and $w=0$ (matter domination) are the two discrete points where the Bardeen scalar potentials $\Psi$ and $\Phi$ have non-dynamical characters \cite{Chu:2016ngc}. In both cases, their sole purpose in describing the tidal forces is only to preserve the causality of $\delta_1 C^i{}_{0j0}$ with respect to their respective background spacetimes.

{\bf De Sitter} \qquad In the de Sitter case ($w = -1$), the background perfect fluid behaves trivially as a cosmological constant, and while the field equations \eqref{PhiPsiRelationship}, \eqref{Vi_Poisson_Cosmology}, and \eqref{Spin-2_WaveEquation_Cosmology} still hold here, the Bardeen scalar $\Psi$ is instead governed by a Poisson-type equation (eq.~(226) of \cite{Chu:2019ndv}),
\begin{align}
(d-2)\vec\nabla^2\Psi&=8\pi G_\text N\big(\rho+(d-1)\mathcal H\Sigma    \big),
\label{Psi_WaveEq_deSitter}
\end{align}
which can be readily solved to yield (eq.~(249) in \cite{Chu:2019ndv})
\begin{align}
\Psi[\eta,\vec x]=\frac{8\pi G_\text N}{d-2}  \int_{\mathbb{R}^{d-1}} \dd^{d-1}\vec x'  \left(G^{\mathrm{(E)}}_d {}^{(\text{a})}T_{00}[\eta,\vec x']+(d-1)\mathcal H[\eta]\, \partial_jD_d  {}^{(\text{a})}T_{0j}[\eta,\vec x']\right),
\label{BardeenPsi_deSitter}
\end{align}
where $\mathcal H = -1/\eta$. This weighted superposition of the matter stress tensor over the constant-time hypersurface clearly indicates the acausal non-radiative nature of $\Psi$ and $\Phi$, which along with $V_i$, do conspire to eliminate the acausal signals from the dynamical spin-2 field $D_{ij}$ within eq.~\eqref{LinearizedWeyl_GaugeInvariant} for $\delta_1 C^i{}_{0j0}$ \cite{Chu:2019ndv}.

Following the same procedure leading up to eq.~\eqref{WeylTensor_RelativisticFluid_General}, the resulting expression of de Sitter $\delta_1 C^{i}{}_{0j0}$ obtained in eqs.~(266) and (267) of \cite{Chu:2019ndv} can in fact be re-written in an analytic compact form for all $d\geq4$, given by
{\allowdisplaybreaks
\begin{align}
\delta_1C^{(\Lambda)i}&{}_{0j0}[\eta,\vec x] = 8\pi G_\text N   \int_{\mathbb{R}^{d-1}} \dd^{d-1}\vec x' \int_{-\infty}^0  \dd\eta' \,\left(\frac{a[\eta']}{a[\eta]}\right)^{\frac {d-2}2}  \Bigg\{  \bigg(\ddot{ G}^{(g|\Lambda,+)}_{d}-(d-3)\mathcal H[\eta]\dot{ G}^{(g|\Lambda,+)}_{d}    \notag\\
&+\frac{(d-2)(d-6)}4\mathcal H[\eta]^2  G^{(g|\Lambda,+)}_{d}\bigg)  \left({}^{(\text{a})}  T_{ij}[\eta',\vec x']+\frac{\delta_{ij}}{d-2}  \left({}^{(\text{a})}  T_{00}[\eta',\vec x'] - {}^{(\text{a})} T_{ll}[\eta',\vec x']  \right) \right) \notag\\
& - 2 a[\eta]^{\frac{d-4}2}  \partial_\eta \left( a[\eta]^{-\frac{d-4}2} \partial_{(i}G^{(V|\Lambda,+)}_{d} \right) {}^{(\text{a})} T_{j)0}[\eta',\vec x']  + \frac{\delta_{ij}}{d-2}  \mathcal H[\eta']   a[\eta]^{\frac{d-4}2} \partial_\eta \left( a[\eta]^{-\frac{d-4}2} G^{(V|\Lambda,+)}_{d} \right)  \notag\\
&  \times \left( (d-3) {}^{(\text{a})} T_{00}[\eta',\vec x']  +  {}^{(\text{a})} T_{ll}[\eta',\vec x'] \right) + \frac{1}{d-2}  \partial_i\partial_j G^{(S|\Lambda,+)}_{d} \left( (d-3){}^{(\text{a})}  T_{00}[\eta',\vec x'] + {}^{(\text{a})}  T_{ll}[\eta',\vec x']  \right) \Bigg\} \notag\\
& + \frac{8\pi G_\text N}{d-2} \left({}^{(\text{a})}T_{ij}[\eta,\vec x]-\frac{\delta_{ij}}{d-1}\left((d-3){}^{(\text{a})}T_{00}[\eta,\vec x]
+2{}^{(\text{a})}T_{ll}[\eta,\vec x]\right)\right) ,
\label{WeylTensor_deSitter_General}
\end{align}}%
where $a[\eta] = -1/(H\eta)$ and the scalar Green's functions $G^{(g|\Lambda,+)}_{d}$, $G^{(V|\Lambda,+)}_{d}$, and $G^{(S|\Lambda,+)}_{d}$ are the de Sitter counterparts of  $G^{(g,+)}_{d}$, $G^{(V,+)}_{d}$, and $G^{(S,+)}_{d}$. Specifically, their even-dimensional forms are given in eqs.~\eqref{GeneralSol_1stGreensFunction_EvenDimensions}, \eqref{GeneralSol_2ndGreensFunction_EvenDimensions}, and \eqref{GeneralSol_3rdGreensFunction_EvenDimensions} with $w$ set to $-1$, namely
{\allowdisplaybreaks
\begin{align}
G^{(g|\Lambda,+)}_{\text{even }d\geq4}[\eta,\eta';R] & = - \Theta[T]  \left(\frac1{2\pi }\frac{\partial}{\partial \overline \sigma} \right)^{\frac{d-2}2} \left( \frac{\Theta[\overline\sigma]}2 P_{\frac{d-2}2}\left[1+\frac{\overline\sigma}{\eta\eta'}\right]\right),   \quad \overline\sigma = \frac{(\eta - \eta')^2 - R^2}2,
\label{1stScalarGreensFunction_Spin2_deSitter} \\
G^{(V|\Lambda,+)}_{\text{even }d\geq4}[\eta,\eta';R] & = - \Theta[T]  \left(\frac1{2\pi }\frac{\partial}{\partial \overline \sigma} \right)^{\frac{d-2}2} \left( \frac{\Theta[\overline\sigma]}2 P_{-\frac{d-2}2}\left[1+\frac{\overline\sigma}{\eta\eta'}\right]\right),
\label{2ndScalarGreensFunction_Spin2_deSitter} \\
G^{(S|\Lambda,+)}_{\text{even }d\geq4}[\eta,\eta';R] & = - \Theta[T]  \left(\frac1{2\pi }\frac{\partial}{\partial \overline \sigma} \right)^{\frac{d-2}2} \left( \frac{\Theta[\overline\sigma]}2 P_{-\frac{d-4}2}\left[1+\frac{\overline\sigma}{\eta\eta'}\right]\right).
\label{3rdScalarGreensFunction_Spin2_deSitter}
\end{align}}%
The derivation of eq.~\eqref{WeylTensor_deSitter_General} then accounts for the result (271) in \cite{Chu:2019ndv}.

It is worth noting that, of all the scalar Green's functions, only the $G^{(g|\Lambda,+)}_{d}$ contains the non-zero tail effect in even dimensions,
\begin{align}
G^{(g|\Lambda,+|\text{tail})}_{\text{even }d\geq4}[\eta,\eta';R] =  - \frac{\Theta[T-R]}{2(2\pi \eta\eta')^{\frac{d-2}2}} ,
\label{1stScalarGreensFunction_Spin2_deSitter_Tail}
\end{align}
obtained by acting with all the differential operators on the Legendre polynomial $P_{\frac{d-2}2}$ in eq.~\eqref{1stScalarGreensFunction_Spin2_deSitter}. However, despite this occurrence, the linearized Weyl tensor $\delta_1C^{(\Lambda)i}{}_{0j0}$ is actually comprised of pure null-cone signals for all even $d\geq4$, as one can readily check by inserting the exact tail function \eqref{1stScalarGreensFunction_Spin2_deSitter_Tail} into eq.~\eqref{WeylTensor_deSitter_General} and inferring that the latter is devoid of tails.

{\bf Matter Domination} \qquad  Next, we turn to the $\delta_1 C^i{}_{0j0}$ in a matter-dominated universe ($w = 0$), which also supplements the causality analysis performed in \cite{Chu:2019ndv}. In this case, the field equations \eqref{PhiPsiRelationship}, \eqref{Vi_Poisson_Cosmology}, and \eqref{Spin-2_WaveEquation_Cosmology} retain the same forms, whereas the Bardeen scalar $\Psi$, according to \cite{Chu:2016ngc}, obeys an ordinary second-order differential equation in time (see eq.~(124) of \cite{Chu:2016ngc}),
\begin{align}
\ddot \Psi + (2d-5) \mathcal H \dot \Psi = 8\pi G_\text N \bigg( \frac{\partial_0 \left(a^{d-2} \Sigma\right)}{(d-2)a^{d-2}} + \mathcal H \dot\Upsilon\bigg),
\label{Psi_WaveEq_MatterDomination}
\end{align}
which is non-dynamical and consistent with the limit of eq.~\eqref{Psi_WaveEq_Cosmology} as $w \to 0$. If the initial value of $\Psi$ and its initial velocity $\dot \Psi$ are both negligible in the asymptotic past, then
\begin{align}
\Psi[\eta,\vec x]= \, & \frac{8\pi G_\text N}{d-2} \int_{\mathbb{R}^{d-1}} \dd^{d-1} \vec x ' \Bigg\{  \frac{(d-1)}2 G^{(\text{E})}_d  \int^{\eta}_0 \dd\eta' \mathcal H[\eta] \mathcal H[\eta'] \left( a[\eta]^{-(d-2)}  \int_{\eta'}^\eta \dd \eta_1 a[\eta_1]^{d-2}  \right)    \left((d-3)  {}^{(\text{a})}T_{00}[\eta',\vec x'] \right. \notag\\
& \left. + {}^{(\text{a})}T_{ll}[\eta',\vec x'] \right)  + G^{(\text{E})}_d {}^{(\text{a})}T_{00}[\eta,\vec x']  + (d-1) \big(\partial_j D_d \big)\mathcal H[\eta] {}^{\text{(a)}}T_{0j}[\eta,\vec x']  \Bigg\},
\label{BardeenPsi_MatterDomination_2}
\end{align}
where the surface terms at spatial infinity and at $\eta = 0$ incurred from integrations by parts have been discarded. The former is justified by the isolated character of the astrophysical system; while the latter by the fact that the boundary term at $\eta \to 0^+$ is actually a homogeneous solution -- whereas what we are after here is the inhomogeneous one. We have massaged $\Psi$ into the form in eq.~\eqref{BardeenPsi_MatterDomination_2} in order to make more transparent the exact cancellation between the acausal portions of the gauge-invariant variables.

Now, inserting the solutions \eqref{BardeenPsi_MatterDomination_2}, \eqref{PhiPsiRelationship_Convolution}, \eqref{VectorMode_RadiationDominated}, and \eqref{Spin2Solution_RadiationDominated} for $w = 0$ into eq.~\eqref{LinearizedWeyl_GaugeInvariant}, and following the same reducing process in the spin-2 sector laid out before, we arrive at, in matter domination,
{\allowdisplaybreaks
\begin{align}
\delta_1C^{(\text M)i}&{}_{0j0}[\eta,\vec x] = 8\pi G_\text N   \int_{\mathbb{R}^{d-1}} \dd^{d-1}\vec x' \int_0^\infty  \dd\eta' \,\left(\frac{a[\eta']}{a[\eta]}\right)^{\frac {d-2}2}  \Bigg\{  \bigg(\ddot{ G}^{(g|\text{M},+)}_{d}-(d-3)\mathcal H[\eta]\dot{ G}^{(g|\text{M},+)}_{d}    \notag\\
&+\frac{(d-2)(2d-7)}4\mathcal H[\eta]^2  G^{(g|\text{M},+)}_{d}\bigg)  \left({}^{(\text{a})}  T_{ij}[\eta',\vec x']+\frac{\delta_{ij}}{d-2}  \left({}^{(\text{a})}  T_{00}[\eta',\vec x'] - {}^{(\text{a})} T_{ll}[\eta',\vec x']  \right) \right) \notag\\
& - 2 a[\eta]^{\frac{d-4}2}  \partial_\eta \left( a[\eta]^{-\frac{d-4}2} \partial_{(i}G^{(V|\text{M},+)}_{d} \right) {}^{(\text{a})} T_{j)0}[\eta',\vec x']  + \frac{\delta_{ij}}{d-2} \mathcal H[\eta']   a[\eta]^{\frac{d-4}2} \partial_\eta \left( a[\eta]^{-\frac{d-4}2} G^{(V|\text{M},+)}_{d} \right)  \notag\\
&  \times \left( (d-3) {}^{(\text{a})} T_{00}[\eta',\vec x']  +  {}^{(\text{a})} T_{ll}[\eta',\vec x'] \right)  + \frac{1}{d-2}\bigg( \frac{2(d-3)}{3d -7} \bigg(  \partial_i\partial_j G^{(S|\text{M},+)}_{d} + \bigg( \frac{d-1}{2(d-3)}\bigg)  \partial_i\partial_j G^{(g|\text{M},+)}_{d}  \bigg) \notag\\
& + \frac{(d-3)(d-1)}2   \mathcal H[\eta] \mathcal H[\eta']    \partial_i\partial_j   Q^{(V|\text{M},+)}_{d}  \bigg)  \left( (d-3){}^{(\text{a})}  T_{00}[\eta',\vec x'] + {}^{(\text{a})}  T_{ll}[\eta',\vec x']  \right) \Bigg\} \notag\\
& + \frac{8\pi G_\text N}{d-2} \bigg({}^{(\text{a})}T_{ij}[\eta,\vec x]-\frac{\delta_{ij}}{d-1}\left((d-3){}^{(\text{a})}T_{00}[\eta,\vec x]
+2{}^{(\text{a})}T_{ll}[\eta,\vec x]\right) + \bigg(\frac{d-3}{3d-7}\bigg) \delta_{ij} \notag\\
&\times \int^{\eta}_0 \dd\eta' \, \bigg(   \mathcal H[\eta] \left(\frac{a[\eta']}{a[\eta]}\right)^{d-2}  - \mathcal H[\eta'] \bigg) \left((d-3)  {}^{(\text{a})}T_{00}[\eta',\vec x] + {}^{(\text{a})}T_{ll}[\eta',\vec x] \right) \bigg) ,
\label{WeylTensor_MatterDomination_General}
\end{align}}%
where $G^{(g|\text{M},+)}_{d}$, $G^{(V|\text{M},+)}_{d}$, and $G^{(S|\text{M},+)}_{d}$ are, again, the three types of massless scalar Green's functions; analogous to those occurring within the Weyl components for other equation-of-states. $Q^{(V|\text{M},+)}_{d}$, in particular, is defined parallel to eq.~\eqref{QV_d},
\begin{align}
Q^{(V|\text{M},+)}_d[\eta,\eta' ;R] & \equiv  a[\eta]^{-\frac {d-2}2}    \int^\eta_{\eta'} \dd \eta_2\,a[\eta_2]^{d-2}\int^{\eta_2}_{\eta'} \dd \eta_1\,a[\eta_1]^{-\frac {d-2}2}    G^{(V|\text{M},+)}_{d}[\eta_1,\eta';R] .
\label{QV_d_MatterDomination}
\end{align}
From eq.~\eqref{WeylTensor_MatterDomination_General} we find that, even though the Bardeen scalars leave some additional local-in-space source terms in the last line, the Weyl components $\delta_1C^{(\text M)i}{}_{0j0}$ are still causally dependent on the matter stress tensor, and may be entirely attributed to the spin-2 gravitons -- as is the case in Minkowski/de Sitter spacetimes whenever the observer is well away from the source. Moreover, like the relativistic-$w$ case in the main text, the integrals involved in eq.~\eqref{QV_d_MatterDomination} cannot be generally performed in a closed form, which in turn prevents $\delta_1C^{(\text M)i}{}_{0j0}$ from being completely analytic. However, within some physical scenarios of cosmological interest, e.g., 4D matter domination, $Q^{(V|\text{M},+)}_d$ can be worked out explicitly.

{\it 4D matter domination} \qquad   In a 4D matter-dominated universe, $a[\eta] = (\eta/\eta_0 )^2$ and $\mathcal H[\eta] = 2/\eta$, the $\delta_1C^{(\text M)i}{}_{0j0}$ components of the linearized Weyl curvature \eqref{WeylTensor_MatterDomination_General} reads
{\allowdisplaybreaks
\begin{align}
\delta_1 C^{(\text{M})i}&{}_{0j0}[\eta,\vec x]
= 8\pi G_\text N   \int_{\mathbb{R}^{3}} \dd^{3}\vec x' \int_0^\infty  \dd\eta' \,\left(\frac{\eta'}{\eta}\right)^2  \Bigg\{  \bigg(\ddot{ G}^{(g|\text{M},+)}_{4}-\frac2\eta\dot{ G}^{(g|\text{M},+)}_{4}    +\frac{2}{ \eta^2} G^{(g|\text{M},+)}_{4}\bigg)  \notag\\
& \times \left({}^{(\text{a})}  T_{ij}[\eta',\vec x']+\frac{\delta_{ij}}{2}  \left({}^{(\text{a})}  T_{00}[\eta',\vec x'] - {}^{(\text{a})} T_{ll}[\eta',\vec x']  \right) \right) - 2 \partial_{(i} \dot G^{(V|\text{M},+)}_{4}  {}^{(\text{a})} T_{j)0}[\eta',\vec x'] \notag\\
&  + \frac{\delta_{ij} }{\eta'}  \dot  G^{(V|\text{M},+)}_{4}   \left(  {}^{(\text{a})} T_{00}[\eta',\vec x']  +  {}^{(\text{a})} T_{ll}[\eta',\vec x'] \right)  + \bigg( \frac15   \bigg( \partial_i\partial_j G^{(S|\text{M},+)}_{4} + \frac32 \partial_i\partial_j  G^{(g|\text{M},+)}_{4}  \bigg) \notag\\
& + \frac3{\eta\eta'}\partial_i\partial_j Q^{(V|\text{M},+)}_{4}  \bigg)    \left( {}^{(\text{a})}  T_{00}[\eta',\vec x'] + {}^{(\text{a})}  T_{ll}[\eta',\vec x']  \right) \Bigg\} \notag\\
&  + 4\pi G_\text N \bigg({}^{(\text{a})}T_{ij}[\eta,\vec x]  -\frac{\delta_{ij}}{3}  \left({}^{(\text{a})}T_{00}[\eta,\vec x]  +  2{}^{(\text{a})}T_{ll}[\eta,\vec x]\right)  + \frac{2\delta_{ij}}5  \int^{\eta}_0 \dd\eta' \, \frac1{\eta'} \bigg(  \left(\frac{\eta'}{\eta}\right)^5  - 1 \bigg) \notag\\
& \times \left(  {}^{(\text{a})}T_{00}[\eta',\vec x] + {}^{(\text{a})}T_{ll}[\eta',\vec x] \right) \bigg) ,
\label{WeylTensor_4DMatterDomination}
\end{align}}%
where the scalar Green's functions $G^{(g|\text{M},+)}_{4}$, $G^{(V|\text{M},+)}_{4}$, and $G^{(S|\text{M},+)}_{4}$ are given by eqs.~\eqref{GeneralSol_1stGreensFunction_EvenDimensions}, \eqref{GeneralSol_2ndGreensFunction_EvenDimensions}, and \eqref{GeneralSol_3rdGreensFunction_EvenDimensions}; with $w = 0$ and $d = 4$:
{\allowdisplaybreaks
\begin{align}
G^{(g|\text{M},+)}_{4}[\eta,\eta';R] & = -\frac{\delta[T-R]}{4\pi R} - \frac{\Theta[T-R]}{4\pi \eta\eta'}  ,
\label{1stScalarGreensFunction_Spin2_4DMatter}\\
G^{(V|\text{M},+)}_{4}[\eta,\eta';R] & = -\frac{\delta[T-R]}{4\pi R}  - \frac{3 \Theta[T-R]}{4\pi\eta\eta'} \left(1 + \frac{(\eta - \eta')^2 - R^2}{2\eta\eta'} \right)  ,
\label{2ndScalarGreensFunction_Spin2_4DMatter}\\
G^{(S|\text{M},+)}_{4}[\eta,\eta';R] & = -\frac{\delta[T-R]}{4\pi R} - \frac{3 \Theta[T-R]}{8\pi\eta\eta'} \left( 5 \left(1+\frac{(\eta - \eta')^2 - R^2}{2\eta\eta'}\right)^2 -1  \right) .
\label{3rdScalarGreensFunction_Spin2_4DMatter}
\end{align}}%
The $Q^{(V|\text{M},+)}_4$ in eq.~\eqref{QV_d_MatterDomination} can now be explicitly evaluated,
\begin{align}
\hspace{-0.47cm} Q^{(V|\text{M},+)}_4[\eta,\eta' ;R]= \frac{\Theta[T-R]}{160\pi \eta^2 \eta'^2} \left( 3 R^4 -  10 R^2 \left(\eta^2 + \eta'^2\right) + 5 \left(3\eta^4 + 2\eta^2\eta'^2 + 3\eta'^4 \right) - \frac{8\big(\eta^5 - \eta'^5 \big) }R   \right) ;
\end{align}
which is a pure tail signal. From these exact expressions, we notice that all relevant Green's functions have non-zero tails. Moreover, according to \eqref{1stScalarGreensFunction_Spin2_4DMatter}, the tail portion of $G^{(g|\text{M},+)}_{4}$ is space independent.

\end{document}